\documentclass[aps,10pt,reprint,groupedaddress,superscriptaddress,twocolumn,longbibliography]{revtex4-2}
\usepackage[usenames,dvipsnames]{xcolor}
\usepackage{amssymb}
\usepackage{graphicx}
\usepackage{amsmath}
\usepackage{dsfont}
\usepackage[bookmarks=true,colorlinks,citecolor=blue,urlcolor=blue]{hyperref}
\usepackage{braket}
\usepackage{babel}
\usepackage{blindtext}
\usepackage{physics}
\usepackage{float}
\usepackage{bm}
\usepackage{float}
\usepackage{tabularx}
\makeatletter
\let\newfloat\newfloat@ltx
\makeatother

\DeclareMathOperator*{\argmax}{arg\,max}
\DeclareMathOperator*{\argmin}{arg\,min}

\usepackage[utf8]{inputenc}
\usepackage[linesnumbered,ruled,vlined]{algorithm2e}
\SetKwInput{KwInput}{Input}                
\SetKwInput{KwOutput}{Output}

\begin{document}

\author{Elisabeth Wybo}
\email{elisabeth.wybo@meetiqm.com}
\affiliation{IQM Quantum Computers, Georg-Brauchle-Ring 23-25, 80992 Munich, Germany}
\author{Jami Rönkkö}
\affiliation{IQM Quantum Computers,
Keilaranta 19, 02150 Espoo, Finland}
\author{Olli Hirviniemi}
\affiliation{IQM Quantum Computers,
Keilaranta 19, 02150 Espoo, Finland}
\author{Jernej Rudi Finžgar}
\affiliation{IQM Quantum Computers, Georg-Brauchle-Ring 23-25, 80992 Munich, Germany}
\author{Martin Leib}
\affiliation{IQM Quantum Computers, Georg-Brauchle-Ring 23-25, 80992 Munich, Germany}

\title{A scalable quantum-enhanced greedy algorithm for maximum independent set problems}

\begin{abstract}
We investigate a hybrid quantum-classical algorithm for solving the Maximum Independent Set (MIS) problem on regular graphs, combining the Quantum Approximate Optimization Algorithm (QAOA) with a minimal degree classical greedy algorithm. The method leverages pre-computed QAOA angles, derived from depth-$p$ QAOA circuits on regular trees, to compute local expectation values and inform sequential greedy decisions that progressively build an independent set. This hybrid approach maintains shallow quantum circuit and avoids instance-specific parameter training, making it well-suited for implementation on current quantum hardware: we have implemented the algorithm on a 20 qubit IQM superconducting device to find independent sets in graphs with thousands of nodes. We perform tensor network simulations to evaluate the performance of the algorithm beyond the reach of current quantum hardware and compare to established classical heuristics. Our results show that even at low depth ($p=4$), the quantum-enhanced greedy method significantly outperforms purely classical greedy baselines as well as more sophisticated approximation algorithms. The modular structure of the algorithm and relatively low quantum resource requirements make it a compelling candidate for scalable, hybrid optimization in the NISQ era and beyond.
\end{abstract}

\maketitle

\section{Introduction}

Quantum computers promise new approaches for solving hard optimization problems~\cite{Abbas2024}. In the current Noisy Intermediate-Scale Quantum (NISQ) era~\cite{Preskill2018}, hybrid quantum-classical algorithms have emerged as leading candidates to achieve quantum utility under realistic hardware constraints. A prominent example is the Quantum Approximate Optimization Algorithm (QAOA)~\cite{Farhi2014,Blekos2024} which is a variational algorithm for solving combinatorial optimization problems that has been the subject of extensive theoretical, numerical and experimental investigation~\cite{Farhi2014,Brandao2018,Wang2018,Basso2021,Farhi2020,Farhi2020a,Farhi2022,Zhou2018,Pagano2020,Harrigan2021,Ebadi2022,Byun2022,Kim2022,Farhi2025}. However, recent theoretical and numerical results provide a mixed outlook on the performance of QAOA. On the one hand, there is evidence that QAOA can outperform the worst-case bound of the best known classical approximation algorithm~\cite{Goemans1995} for solving the MaxCut problem when the QAOA circuit depth $p$ is sufficiently large but independent of the problem size~\cite{Brandao2018,Crooks2018,Wurtz2021,Farhi2025}. On the other hand, there are also rigorous indications that constant-depth QAOA has limitations~\cite{Farhi2020,Farhi2020a,Chou2021,Barak2021,Anshu2022,Chen2023}. For instance it provably cannot solve the Maximum Independent Set (MIS) problem to (near) optimality due to its locality~\cite{Farhi2020,Farhi2020a}. This implies that unbounded depth is a necessary, but not sufficient, condition for reaching (near) optimal solutions. However, scaling QAOA to larger problem sizes in practice remains challenging because current NISQ devices can only support shallow circuits before noise dominates~\cite{Franca2020,Wang2021,DePalma2022,Gonzlez_Garca_2022}. 

A natural response to these limitations is to reconsider the role of the quantum algorithm itself. Rather than viewing QAOA as a standalone optimizer that must fully solve the problem, one can instead use it as a subroutine that enhances classical heuristics~\cite{brady2024,Dupont2024,Dupont2023a,Finzgar2024}. In this work, we adopt this perspective and study a quantum-enhanced greedy algorithm for the MIS problem~\cite{brady2024}. Greedy algorithms are attractive because of their simplicity and scalability~\cite{Frieze1990,Frieze1992,Halldorsson1992,Halldorsson1998,Wormald1999}. However, their decisions are based on limited information available at each step and can therefore miss important but intricate structure in the problem. The central idea of the hybrid approach is to improve these decisions using expectation values from QAOA. Specifically, at each iteration, a shallow QAOA circuit is executed on a residual problem graph to estimate expectation values of local observables, which are then used to guide the next greedy step~\cite{brady2024,Finzgar2024}. In this way, the quantum processor provides structured guidance derived from an entangled quantum state, while the classical algorithm retains control over the iterative reduction of the problem.

The central contribution of this work is an efficient and scalable version of the quantum-enhanced greedy algorithm that eliminates the need for variational parameter optimization thus retaining an overall linear-time complexity. Optimizing variational QAOA angles is widely recognized as a major bottleneck particularly as system size increases, due to highly non-convex energy landscapes and the onset of barren plateaus~\cite{Wang2021,Cerezo2021,Larocca2025}. We bypass this challenge entirely by employing fixed-angle QAOA circuits with parameters derived from regular tree models (Bethe lattice like structures), which are well suited to problems on large $d$-regular graphs that resemble trees locally~\cite{Streif2020,Wurtz2020,Basso2021,Wybo2025}. This enables a plug-and-play use of QAOA without instance-specific tuning. Crucially, for fixed QAOA depth $p$ and fixed degree $d$, each greedy update requires the evaluation of a constant number of local expectation values and a constant amount of classical post-processing. As a result, the total run time of the quantum-enhanced greedy algorithm scales as $O(N)$ with $N$ the number of nodes of the problem graph.

To benchmark the method beyond current hardware capabilities, we compute the required QAOA expectation values using tensor-network contractions, allowing us to study the shallow-depth regime ($p \leq 4$) and elucidate the exponential cost in $p$ induced by light-cone growth. In numerical simulations on random $3$-regular graphs, the quantum-enhanced greedy algorithm consistently outperforms the classical greedy baseline. We observe the same behavior experimentally by implementing the algorithm at depths $p=2,3$ on the IQM Garnet quantum processor~\cite{abdurakhimov2024}. 
Finally, we compare the hybrid approach to the linear-time prioritized search algorithm of Marino \emph{et al.}~\cite{Marino2020} and find improved performance up to $N \approx 5000$ nodes for $p=4$. This is significant because Marino’s algorithm is a recent linear-time prioritized search algorithm  specifically tailored to random regular graphs, yet our approach already outperforms it using only shallow circuits to estimate local expectation values. By using QAOA expectation values to guide a classical greedy search, our approach partially mitigates current hardware limitations that preclude the much deeper circuits required by a standalone quantum algorithm. Therefore, our results suggest that classical optimization can benefit substantially from incorporating quantum components even in the near term. The present work illustrates how modest quantum resources integrated into classical strategies can enhance heuristic optimization and lead to better outcomes than either method alone.

The remainder of the paper is organized as follows. Section~\ref{sec:model} introduces the MIS problem and its formulation as an Ising model. Section~\ref{sec:main} discusses the quantum-enhanced greedy algorithm: \ref{subsec:qaoa} reviews background about QAOA, \ref{subsec:algo} outlines the algorithm, \ref{subsec:tns} contains the tensor-network simulation results, \ref{subsec:hw_complexity} discusses the complexity of computing local expectation values, \ref{subsec:implementation} discusses an experimental implementation of the algorithm on current quantum hardware and \ref{subsec:noise} analyzes the effects of noise. Section~\ref{sec:discussion} presents a comparison of our hybrid greedy approach with classical algorithms, highlighting relative performance and scaling. We conclude in Section~\ref{sec:concl} with a summary of our findings and an outlook on future improvements and generalizations of the QAOA-greedy paradigm.

\begin{figure}
    \centering
    \includegraphics[width=0.49\textwidth]{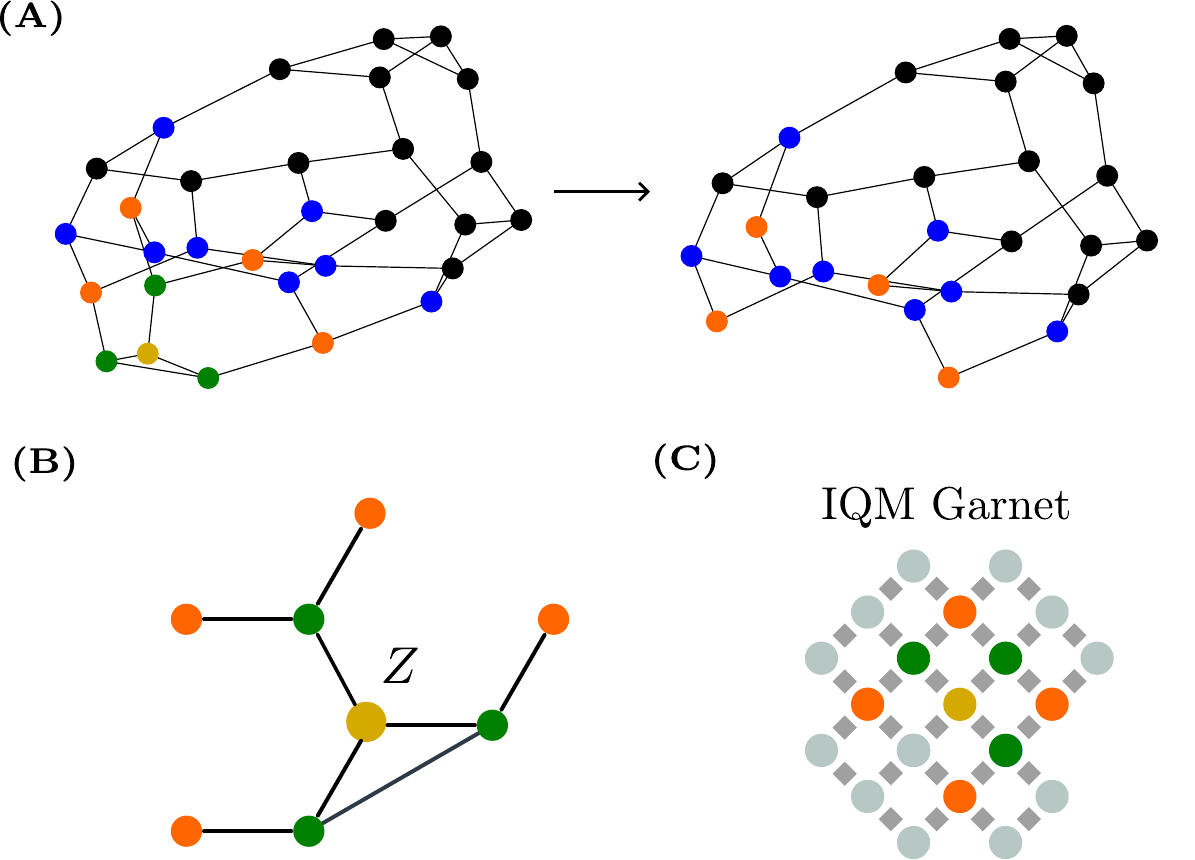}
    \caption{(A) First iteration of the hybrid quantum-classical greedy algorithm for QAOA depth $p=2$ on a 3-regular graph of $N=30$ nodes. The yellow node is selected as a node that has maximal $\ev{Z}_{p=2}$. The nodes of its $p=1$ subgraph are shown in green, the additional nodes in its $p=2$ subgraph in orange. The yellow node is added to the independent set and its direct neighbours (i.e. the nodes in the $p=1$ subgraph, the green nodes) are deleted from the graph. This deletion affects the expectation value of the remaining blue and orange nodes in the graph which need to be recomputed. (B) The $p=2$ subgraph that determines the expectation value. (C) The expectation values corresponding to the different subgraphs are computed using a superconducting 20 qubit IQM device.   }
    \label{fig:sketch}
\end{figure}

\section{Maximum independent set problem and classical algorithms} \label{sec:model}

The Maximum Independent Set (MIS) problem is a fundamental combinatorial optimization problem in graph theory. Given a graph $G=(V,E)$, an \emph{independent} set is a subset of vertices such that no two vertices in the set are adjacent, i.e. $\forall i,j \in \mathcal{I}: ij\notin E$, where $E$ is the edge set associated with $G$. The MIS problem asks for an independent set of maximum cardinality. This problem is NP-hard in general graphs and remains hard to approximate within any constant factor unless $\mathrm{P}=\mathrm{NP}$~\cite{GareyJohnson1979,Hastad1999}.
This problem, and related problems such as the Maximum Clique and Minimum Vertex Cover, have many real-world applications in various domains~\cite{Wurtz2024}, including logistics~\cite{Kieritz2010}, computer graphics~\cite{Sander2008} and computational biology~\cite{Butenko2006,Cheng2008}. The MIS problem can be cast into finding the ground state of an Ising model. The solution to the problem is a basis state $\ket{\bm{z}}$ (or a superposition thereof) that minimizes the energy $H_{\lambda}(\bm{z})=\ev{H_{\lambda}}{\bm{z}}$. The Ising Hamiltonian associated with solving the MIS problem is given by
\begin{equation}\label{eq:MIS}
H_{\lambda} = \lambda \sum_{ij \in E} N_i N_j - \sum_{i\in V} N_i 
\end{equation}
with $\lambda\geq 1$. Here the `number operator' is defined as $N_i = \frac{Z_i + 1}{2}$ in terms of the Pauli $Z$ operator associated with the $i$th variable, and acts as $\ev{N_i}{z_i} \in \{ 0,1\}$. In this notation the second term in Eq.~\eqref{eq:MIS} is counting the number of vertices in the set, while the first term ensures the independence constraint by penalizing the inclusion of adjacent vertices in the set. In terms of Pauli $Z$ operators, the problem Hamiltonian becomes 
\begin{equation} \label{eq:MIS_z}
H_{\lambda} = \frac{\lambda}{4} \sum_{ij \in E} Z_i Z_j  + \sum_{i\in V} \frac{\lambda d_i -2}{4}  Z_i +  \sum_{i\in V} \frac{\lambda d_i -4}{8},
\end{equation}
where $d_i$ denotes the degree (i.e. the number of neighbors) of node $i$. 
 
 A simple and efficient classical algorithm that is guaranteed to find independent sets is the minimal greedy search algorithm. This algorithm takes the following steps: (i) Randomly select a vertex from the subset of vertices of $G$ that have the lowest degree. (ii) Add this vertex to the independent set, and delete all its neighbors from $G$. (iii) Repeat (i) and (ii) until the graph $G$ is empty (i.e. has no vertices left). This algorithm thus makes locally optimal choices under the assumption that nodes with minimal degree have priority to be included in the independent set. This greedy algorithm is guaranteed to find independent sets $\mathcal{I}$ with a worst-case approximation ratio~\cite{Hallorsson1997}
 \begin{equation} \label{eq:greedy_perf}
     \alpha_{\textrm{Greedy}} = |\mathcal{I}|/|\mathcal{I}_{\textrm{max}}| \geq 3/(d+2)
 \end{equation}
 for graphs with bounded degree $d$, where $\mathcal{I}_{\textrm{max}}$ is the size of the MIS. However, greedy search is not the best efficient classical algorithm. In particular for regular graphs, where every node has fixed degree $d$, an improved linear-time prioritized search algorithm has been developed in Ref.~\cite{Marino2020}. In fact, this algorithm improved upon the previously known independence ratios $r=|\mathcal{I}|/|V|$ for random $d$-regular graphs~\cite{Csoka2015}, and is therefore the State-Of-The-Art (SOTA) classically efficient $O(N)$ algorithm. The idea behind this algorithm is to defer decisions of which nodes to add to $\mathcal{I}$ by contracting nodes into larger structures called `virtual nodes'. Crucially, it is such a contraction sequence, in contrast to the random deletion sequence of the minimal greedy algorithm, that explains the better performance of this algorithm. 
 
 For explicit comparison, in the case of large $d=3$ regular graphs the greedy algorithm achieves an average independence ratio $\mathbb{E}_G(r) =  6\log(3/2)-2 \approx 0.432\dots$~\cite{Wormald1999} (Eq.~\eqref{eq:greedy_perf} is the worst case bound), while the linear-prioritized search algorithm of Ref.~\cite{Marino2020} achieves $\mathbb{E}_G(r) \approx  0.445\dots$. In the remainder of the paper, we investigate a hybrid quantum-classical approach for 3-regular graphs and compare explicitly to the classical approaches.

\section{Quantum-enhanced greedy algorithm} \label{sec:main}
In this section, we will explain the quantum-enhanced greedy method. The central idea of the algorithm is to boost the performance of the classical greedy algorithm using expectation values obtained from a QAOA state. Such a hybrid scheme for MIS has been proposed in a general setting in Ref.~\cite{brady2024}. In this work, we tailor the method to $3$-regular graphs and combine it with the fixed-angle QAOA of Ref.~\cite{Wybo2025}, allowing for efficient large-scale simulations without the need for parameter optimization loops. This reduces the resources required both for the classical analyses as well as for implementations on quantum hardware. As the quantum greedy algorithm takes decisions based on expectation values from a QAOA circuit, we will start by briefly reviewing QAOA.

\begin{figure}
    \centering
    \includegraphics[width=0.49\textwidth]{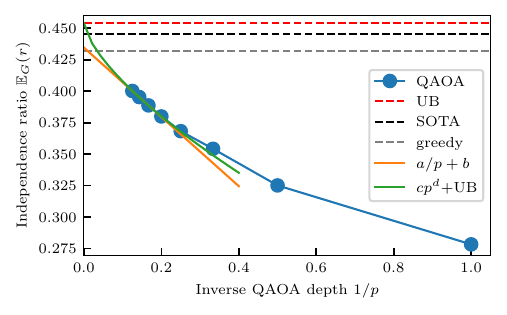}
    \caption{\textbf{Performance of QAOA in solving the MIS problem on 3-regular graphs} as a function of $1/p$ in the limit $N\rightarrow\infty$. We have performed two polynomial fits $a/p+b$ and $cp^d$ with corresponding fitting parameters $(a,b)=(-0.275,0.435)$ and $(c,d)=(-0.222,0.683)$ to indicate that these results suggest that QAOA needs significant depth to match the minimal greedy algorithm~\cite{Frieze1994} (grey dashed line). We also compare to the performance of the linear-time algorithm of Ref.~\cite{Marino2020} (SOTA, black dashed line) and an upper bound on the size of the independent set (UB, red dashed line)~\cite{McKay1987,Balogh2017}.  }
    \label{fig:qaoa_perf}
\end{figure}

\subsection{Quantum approximate optimization algorithm} \label{subsec:qaoa}
The Quantum Approximate Optimization Algorithm (QAOA) is a variational quantum algorithm designed to approximately solve combinatorial optimization problems. Such problems can be expressed as diagonal Hamiltonians like the MIS Hamiltonian in Eq.~\eqref{eq:MIS_z}. QAOA, as introduced in Ref.~\cite{Farhi2014}, defines a depth-$p$ circuit ansatz composed of alternating applications of unitaries generated by the problem Hamiltonian $H$ and mixing Hamiltonian $B$
\begin{equation} \label{eq:qaoa_ansatz}
\ket{\bm{\gamma},\bm{\beta}}
= e^{-i\beta_p B}e^{-i\gamma_p H}
\cdots
e^{-i\beta_1 B} e^{-i\gamma_1 H}
\ket{+},
\end{equation}
where $B = \sum_i X_i$ and the initial state $\ket{+} = 2^{-N/2}\sum_{\bm{z}}\ket{\bm{z}}$.

The task is to determine parameters $\bm{\beta}=(\beta_1,\dots,\beta_p)$ and $\bm{\gamma}=(\gamma_1,\dots,\gamma_p)$ that minimize the variational energy,
\begin{equation} \label{eq:qaoa_var}
(\bm{\gamma}^{\star},\bm{\beta}^{\star})
= \argmin_{\bm{\gamma},\bm{\beta}}
\ev{H}{\bm{\gamma},\bm{\beta}}.
\end{equation}
Sampling $\bm{z} \in \{-1,1\}^N$ from the optimized state $\ket{\bm{\gamma}^{\star},\bm{\beta}^{\star}}$ then yields configurations with low energy $H(\bm{z})$, which are approximate solutions to the underlying problem.
While QAOA can be viewed as a discretized version of adiabatic evolution, the adiabatic theorem does not imply that QAOA with finite depth can reach the optimum. Rather, in the limit of  $p\rightarrow\infty$, the adiabatic theorem and vanishing Trotter errors, guarantee convergence to the ground state of the problem Hamiltonian.

\begin{algorithm}[t!]
   \caption{Quantum-enhanced greedy search}
   \label{alg:q_greedy}
   \KwInput{$G(V,E)$, $\bm{\beta}^{\star}$, $\bm{\gamma}^{\star}$;}
   \KwOutput{$\mathcal{I}$;}
    $\mathcal{I} \gets \emptyset $;\\
    $N\gets \emptyset $;\\
    Prepare QAOA state $\ket{\bm{\gamma}^{\star},\bm{\beta}^{\star}}$ based on $G$; \\
        Construct a dictionary of local $Z$-measurements $S = \{ j: \ev{Z_j}{\bm{\gamma}^{\star},\bm{\beta}^{\star}}, \forall j \in V \}$; \\
    \While{$G$ is not empty}{
        Recompute $ j: \ev{Z_j}{\bm{\gamma}^{\star},\bm{\beta}^{\star}}, \forall j \in N $; \\
        Update $S$; \\
        Select node $i \in V$ with the largest exp. val.
        $i\gets \argmax_{j\in V}\ev{Z_j}{\bm{\gamma}^{\star},\bm{\beta}^{\star}}$;\\
        Add $i$ to $\mathcal{I}$;\\
     $N \gets \{\, j \in V : \mathrm{dist}_G(i,j) \le p+1 \,\}$;\\
        Delete $i$ and its neighbors from $G$; \\
        }
    return $\mathcal{I}$
\end{algorithm}

A key feature of QAOA on large random $d$-regular graphs is that the optimal variational parameters are independent of the problem size and of the specific instance, but depend exclusively on the degree $d$~\cite{Basso2021,Wybo2025}. This is a direct consequence of the locality of depth-$p$ QAOA. The expectation value of each local term in the Hamiltonian is determined solely by the structure of the problem graph within its depth-$p$ light cone (see Fig.~\ref{fig:sketch}{(A)}). For large random $d$-regular graphs, the light-cone induced subgraphs are regular trees with high probability, and deviations due to short cycles become increasingly rare as the graph size grows. Therefore, the local neighborhoods of all vertices converge to $d$-regular trees. This implies that the QAOA energy~\eqref{eq:qaoa_var} can be evaluated --- exactly in the infinite-size limit and to very good approximation for large finite graphs --- by considering only tree structures. Optimal (or near-optimal) QAOA parameters can therefore be computed once on the corresponding regular tree, and then applied uniformly to all problem instances.
The resulting ``tree'' angles for MIS on $d=3$ regular graphs were calculated in Ref.~\cite{Wybo2025} and explicitly given in Appendix~B. In this work, we use these precomputed tree angles as fixed QAOA parameters within our quantum greedy algorithm, thereby avoiding any parameter-optimization loops. 

The independence ratios obtained from the tree-based $N\to\infty$ analysis of QAOA are shown in Fig.~\ref{fig:qaoa_perf}. We compare these values with the performance of the greedy baseline, the state-of-the-art classical algorithm, and an upper bound on optimality. For the range of QAOA depths considered, we find that the guaranteed stand-alone performance of QAOA remains far below the greedy algorithm. Moreover, extrapolating polynomial fits to the data suggests that $p\gtrsim 50$ is required only to match the asymptotic performance of the minimal degree greedy algorithm. This comparatively low performance on MIS motivates the need for more tailored approaches, such as the quantum-enhanced greedy algorithm described below.

\begin{figure}[t]
    \centering
    \includegraphics[width=0.49\textwidth]{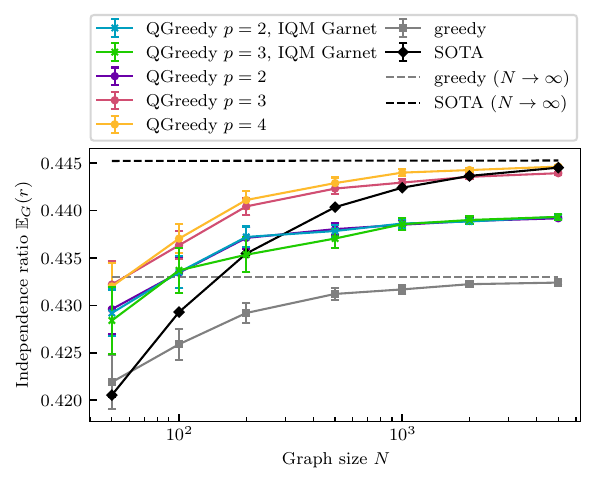}
    \caption{\textbf{Independence ratios obtained by the quantum greedy algorithm} compared with the independence ratios found by the greedy and linear-prioritized search algorithm for $d=3$ regular graphs. The dotted line corresponds to $r_{\infty}=0.445330$ which is the lower bound on the independence ratio in the limit $N\rightarrow \infty$ given in Ref.~\cite{Marino2020}. The error bars show three times the standard error of the mean originating from the average over graph instances. There are $200$ graph instances for QGreedy with each $p$, except for $p=3$ case with IQM Garnet, which has $100$ instances. The considered graph sizes are $N=50,100,200,500,1000,2000,5000$.}
    \label{fig:result}
\end{figure}

\subsection{Quantum enhanced greedy algorithm}\label{subsec:algo}
Let us now describe the quantum-enhanced greedy algorithm. The key idea is to keep the classical greedy reduction as the algorithmic backbone and use a quantum subroutine to refine the selection rule at each step. In the purely classical baseline, a node $i \in V$ with minimal degree is selected uniformly at random and added to the independent set, after which its neighbors are deleted. The quantum enhanced version replaces this random choice by evaluating local QAOA expectation values $\ev{Z_i}{\bm{\gamma}^{\star},\bm{\beta}^{\star}}\equiv\ev{Z_i}_p, \forall i \in V$ and by selecting the node that has the highest expectation value~\cite{brady2024}. This node also has the highest value of $\ev{N_i}{\bm{\gamma}^{\star},\bm{\beta}^{\star}}\in [0,1]$ which can be interpreted as having the highest probability to be in the independent set. Since the QAOA angles are fixed, the value of $\ev{Z_i}_p \in [-1,1]$ depends only on the topology of the (rooted) light-cone induced subgraph $\tilde{G}_i \subseteq G$ that contains the distance $p$ neighbourhood of node $i$. The algorithm is described in Algorithm~\ref{alg:q_greedy}.

At QAOA depth $p=1$, the method reduces to the classical greedy baseline. This can been seen from the analytical expressions of the local expectation values at $p=1$ (see e.g. Ref.~\cite{Ozaeta2020})
\begin{equation} \label{eq:p1_localterm}
    \ev{Z_i}{\gamma_1^{\star},\beta_1^{\star}} = \sin(2\beta_1^{\star}) \sin(2\gamma_1^{\star} h_i) \left[ \cos(2\gamma_1^{\star}) \right]^{d_i}.
\end{equation}
Here, $h_i = (\lambda d_i -2 )/4$ (see Eq.~\eqref{eq:MIS_z}) denotes the local field associated to the $i$-th node. The optimal $2\gamma_1^{\star}$ and $2\beta_1^{\star}$ found in Ref.~\cite{Wybo2025} lie in the first and fourth quadrant (or vice-versa) of the circle, respectively. As we take these angles to be fixed, it follows that the expectation value~\eqref{eq:p1_localterm} is maximal when $d_i$ is minimal, thus recovering the classical minimal-degree selection rule.

For $p>1$, the QAOA circuit introduces longer-range correlations across each light cone, and therefore the expectation values $\ev{Z_i}_p$ incorporate richer structure. These expectation values allow the quantum-enhanced greedy algorithm to prioritize nodes more effectively by breaking ties that would otherwise be broken uniformly at random by the bare greedy algorithm. This allows our approach to improve upon the classical baseline as we will demonstrate in simulation and experiment in later sections. 



When running the algorithm as described above on a $d$-regular graph, only the original graph is guaranteed to be $d$-regular. During the algorithm nodes are removed from the graph causing the average degree $1/|V| \sum_{i\in V}d_i$ to decrease below $d$. Yet, we still use the fixed tree angles for $d$-regular graphs throughout the algorithm to avoid the computational overhead of variational parameter optimization. This is justified because the local subgraphs retain key structural features of the original $d$-regular graph, particularly in early iterations of the algorithm where most of the crucial steps for improving on greedy are made. Moreover, we find empirically that this choice leads to a substantial improvement upon the classical greedy baseline.

Due to the locality of QAOA, a full recomputation of the expectation values is not always necessary at every iteration. Indeed, if we remove the node $i$ and its neighbors from the graph, this change will only affect nodes that are within a distance $p+1$ from $i$, as illustrated in Fig.~\ref{fig:sketch}. Hence, only the expectation values of the affected nodes need to be recomputed. Moreover, the expectation value $\ev{Z_i}_p$ is equal for nodes with isomorphic light cone subgraphs (because we consider unweighted MIS problems). The number of possible subgraphs grows at least exponentially with $p$, which makes a full enumeration infeasible for $p>3$. However, we can still implement a simple cache, where the expectation value for each non-isomorphic subgraph is stored in a dictionary that can be re-used if needed, thus increasing the efficiency of the algorithm in practice. In Table~\ref{tab:counts} we show the number of all non-isomorphic subgraph topologies of the rooted subgraphs $\tilde{G}_i$ for the lowest values of $p$. However, we stress that, although the space of all rooted light-cone topologies grows (super-) exponentially with $p$, the greedy step on a fixed instance only ever compares at most $N$ local values. Thus, what matters in practice is not discriminating all possible subgraph topologies, but resolving the exponentially smaller set of distinct light cones topologies that actually occur in the given graph and, in particular, the difference between the top few expectation values that determine the greedy choice.

For fixed $p$ the cost of computing $\ev{Z_i}_p$ only depends on $p$ (and not on $N$). Therefore, for \emph{fixed} QAOA depth $p$, the run time of the quantum greedy algorithm scales as $O(N)$ and therefore has the same overall scaling as the bare greedy algorithm. 
However, it is known that the bare QAOA algorithm has limitations when $p$ is chosen to be in $o(\log(N))$, i.e. when QAOA does not `see' the whole graph~\cite{Farhi2020a}. This suggests that $p$ will need to grow at least as $\log(N)$ to find near-optimal solutions. This `see-the-whole-graph' criterion would increase the run time scaling of the algorithm to $\Omega(N\log(N))$. 

\begin{table}[t!]
\centering
\begin{tabular}{c r r r}
\hline\hline
$p$ & Total & Trees & Non-Trees \\
\hline
1 & 4     & 4     & 0      \\
2 & 75    & 20    & 55     \\
3 & 44502 & 286   & 44216  \\
\hline\hline
\end{tabular}
\caption{Total number of non-isomorphic subgraphs (tree and non-tree) $\tilde{G}\subseteq G$ for small values of \( p \) that can be encountered by the quantum greedy algorithm for solving MIS on 3-regular graphs.}
\label{tab:counts}
\end{table}

The quantum-enhanced greedy algorithm offers two clear advantages over vanilla QAOA. First, it is inherently constructive: by design, it always produces a valid independent set. This property does not hold for vanilla QAOA~\cite{Farhi2020a}, where samples drawn from the output distribution can violate independence constraints at finite circuit depth. Second, the quantum greedy algorithm exhibits a degree of intrinsic robustness to noise. Indeed, because the quantum subroutine is used only to inform individual greedy decisions within an otherwise classical procedure, moderate noise degrades the quality of the advice but cannot cause the algorithm to produce an invalid independent set. A detailed discussion and quantitative analysis of this robustness are deferred to Sec.~\ref{subsec:noise}. Together, these features illustrate how embedding quantum subroutines within a larger hybrid workflow can address some of the limitations of standalone quantum algorithms.

\begin{figure*}
    \centering
    \includegraphics[width=1.02\textwidth]{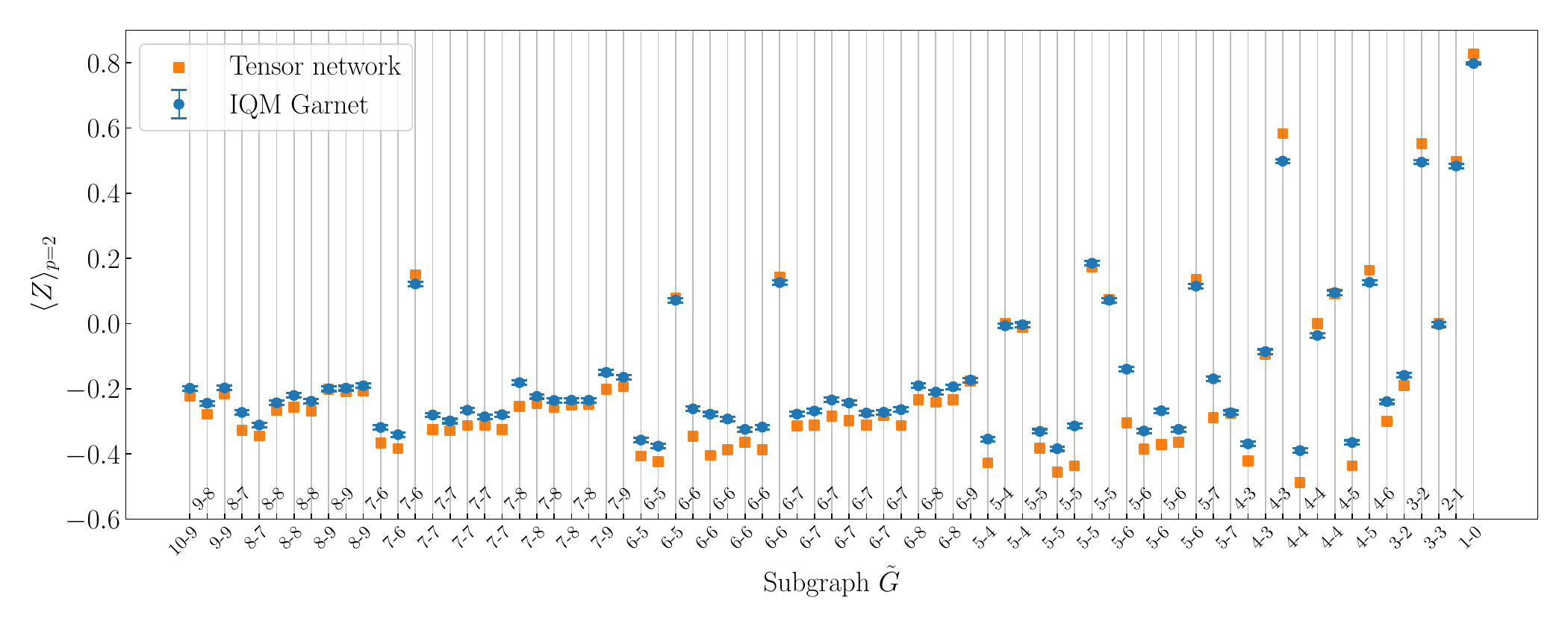}
    \caption{\textbf{Expectation values corresponding to all 75 possible subgraphs} that can occur during the quantum-enhanced greedy search at depth $p=2$. The labels on x-axis denote the number of nodes and edges in each subgraph. Expectation values are displayed in order of descending number of nodes and ascending number of edges. The orange markers show the ideal results from tensor network simulations and blue markers show the corresponding results from the IQM Garnet quantum device. These expectation values are used to decide which node is added to the independent set: from all the subgraphs present in each iteration of the algorithm, the one with largest $\ev{Z_i}$ is chosen. The error bars mark standard deviation from $1000$ bootstrapped samples. They lie fully within the markers, signifying good statistical accuracy thanks to $20 000$ quantum circuit measurements for each data point.} 
    \label{fig:z_exp}
\end{figure*}

\subsection{Tensor-network simulations} \label{subsec:tns}


In this section, we present classical tensor-network simulations of the QAOA light-cone expectation values and analyze the associated computational cost. This provides a performance benchmark for the hybrid algorithm beyond current hardware. Moreover, by quantifying the classical resources required to compute the same local quantities, we can directly assess the efficiency gains of estimating them on quantum hardware.

For the tensor-network simulations we use the publicly available \texttt{QUIMB} package~\cite{Gray2018quimb}. We have performed simulations up to $p=4$ for $d=3$ and graph sizes up to $N=5000$ (see Fig.~\ref{fig:result}). We sampled 200 graph instances for each problem size, using the same instance set across all depths $p$, and averaged the resulting independence ratios over these instances. The error bars show the standard error of the mean. In Fig.~\ref{fig:z_exp} we show the expectation values of all of the 75 possible subgraphs for $p=2$ (see Table~\ref{tab:counts}).

 In the tensor-network picture every gate in the QAOA circuit is represented as a tensor. To evaluate $\ev{Z_i}_p=\Tr(Z_i\dyad{\bm{\gamma}^{\star},\bm{\beta}^{\star}})$, we construct the light cone of the operator $Z_i$. This means that we select only the tensors inside the tensor-network representation of $\dyad{\bm{\gamma}^{\star},\bm{\beta}^{\star}}$ that influence this expectation value (see Fig.~\ref{fig:sketch}). We will denote the corresponding circuit graph as $\tilde{G}_i^C$~\cite{Shi2008}. The cost of computing the expectation values on the light cone depends on the chosen contraction path. In the worst case, this can be as expensive as state vector simulation for which the cost is $O(\exp(|\tilde{V}_i|))$ where $|\tilde{V}_i|$ is the number of vertices in the subgraph $\tilde{G_i}$. However, as shown in Ref.~\cite{Shi2008} the contraction path can be optimized such that the contraction cost scales as $\exp(O(\mathrm{tw}(\tilde{G}^C_i)))$. 
 If $\tilde{G}_i$ is a tree, it follows that $\mathrm{tw}(\tilde{G}^C_i)\in O(p)$, implying that the overall contraction cost scales as $\exp(O(p))$. Such scaling can be understood from bounds on the treewidth of the circuit graph, which is constructed as the Cartesian product of a tree and a path graph. From the definition of treewidth, the treewidth of the circuit graph can be upper bounded by $4p-1$. From Ref.~\cite{Wood2011} it can be lower bounded by $2p-1$. Taken together, this yields the $\exp(O(p))$ scaling of the tensor-network simulation cost. 
 In contrast, if $\tilde{G}_i$ is not a tree, then $\mathrm{tw}(\tilde{G}^C_i) \in O(\exp(|\tilde{V}_i|))$. The number of nodes in the subgraph $\tilde{G}_i$ is generally (if not limited by the actual problem size) scaling as $O(\exp(p))$ and is upper bounded by the number of nodes in the depth-$p$ $d$-regular tree
\begin{equation} \label{eq:nodes_in_tree}
    N_{\textrm{tree}}(p,d) = 1+d\frac{(d-1)^{p}-1}{d-2}.
\end{equation}
 Therefore, if $\tilde{G}_i$ is not a tree, the contraction cost of the circuit graph may scale as a double exponential in $p$ in the worst case. We conclude that the tree is the `cheapest' structure to contract, with exponential scaling in $p$, although it does contain the most nodes. 
 

\subsection{Complexity of estimating expectation values on quantum hardware} 
\label{subsec:hw_complexity}

The cost of implementing QAOA subgraphs on (noiseless) quantum computers is determined by the routing overheads required to execute the circuit on locally connected hardware~\cite{Kotil2023}. In particular, Refs.~\cite{Schnorr1986,Steiger2019} show that any quantum circuit of depth $p$ on $|\tilde{V}_i|$ qubits can be implemented on a square lattice of qubits in depth $O(p |\tilde{V}_i|^{1/2})$. 
Therefore, since $|\tilde V_i|$ grows as $O(\exp(p))$, the expected speedup over exact classical simulation ranges from polynomial for tree-like light cones, to exponential when the induced circuit graph has treewidth $O(|\tilde V_i|)$.
We summarize the computational cost of evaluating QAOA expectation values as a function of the circuit depth $p$ in Table~\ref{tab:cost}. A key observation is that the size of the light-cone subgraph $\tilde{G}_i$ for fixed $p$ does not depend on the problem size $N$. Consequently, for fixed depth $p$, the cost of evaluating local expectation values remains constant in $N$, and the overall quantum-enhanced greedy algorithm scales linearly.

Overcoming the intrinsic locality of QAOA requires the light cones to span the entire graph. As shown in Ref.~\cite{Farhi2020}, this necessitates circuit depths scaling $p \in \Omega(\log N)$ on bounded-degree graphs. Substituting $p=\log N$ into the cost scaling of computing $\ev{Z_i}_p$ Table~\ref{tab:cost}, the run time of the quantum-enhanced greedy algorithm would increase to $O(N^{3/2})$ assuming the locally connected hardware.

An additional consideration is the additive precision up to which the $\langle Z_i\rangle_p$'s must be estimated in order to carry out the greedy step reliably. For a fixed instance, the algorithm compares at most $N$ such values, and the relevant quantity is the separation $\Delta_p$ between the largest and second-largest expectation values. Each empirical estimate $\widehat z_i$ that is obtained from $M$ independent projective measurements of a $\pm 1$-valued observable and has variance $\mathrm{Var}(\widehat z_i)\le 1/M$. By Hoeffding's inequality~\cite{Hoeffding1963}, the probability that any given estimate deviates from its expected value by more than $\delta$ is upper bounded by $\exp(-2M\delta^2)$. Requiring that all $N$ estimates simultaneously lie within $\Delta_p/2$ of their true values, and applying a union bound over the $N$ candidates gives the sufficient condition $N \exp^{-M\Delta_p^2/2} \leq \epsilon$, which can be rearranged to 
\begin{equation}
    M \geq \frac{\log(N/\epsilon)}{\Delta_p^2}. 
\end{equation}
In practice, however, $\Delta_p$ is not known a priori and may be very small or even vanish. This motivates introducing a depth-dependent cutoff $\delta_p$ which sets the number of shots and effectively treats nodes with $|\ev{Z_i}-\ev{Z_j}| \leq \delta_p$ as degenerate. This cutoff could for instance be chosen to be the smallest change of $\langle Z_i\rangle_p$ obtained by adding a single edge to the depth-$p$ single root tree corresponding to closing the simplest resolvable cycle.

To assess the relevance of the quantum subroutine that computes local expectation values, it is essential that its output cannot be efficiently reproduced by classical means. Therefore, we conclude this section by discussing the classical hardness of computing local expectation values of QAOA circuits. Under mild assumptions, QAOA circuits are universal and can approximate arbitrary quantum computations~\cite{Brylinski2001,Morales2019}. Consequently, evaluating their expectation values is expected to be $\mathrm{BQP}$-hard~\cite{Lloyd2018}. Moreover, recent work~\cite{Wang2025} has formally established the relationship between the classical hardness of evaluating QAOA expectation values and the tensor-network structure of the circuit graph $\tilde{G}^C_i$. In particular, it is shown that for any depth $p>1$ computing (or even approximating up to additive precision) expectation values from a QAOA circuit is NP-hard. Their reduction proceeds by encoding the solution of an NP-hard optimization problem into a coefficient of the Laurent polynomial obtained from $\tilde{G}^C_i$ and provides a complexity-theoretic formalization of the tensor-network intuition, i.e. once the circuit graph has grown to large treewidth the associated tensor network cannot be efficiently contracted, and the expectation value becomes intractable. In summary, the classical hardness of obtaining the expectation values used in the quantum-greedy algorithm follows directly from the exponential tensor-network contraction cost set by its light-cone subgraph $\tilde{G}_i$ such that the quantum speedups for estimating expectation values that are listed in Table~\ref{tab:cost} are expected to hold in general.

\begin{table}[t!]
\begin{center}
\begin{tabular}{lll}
\hline\hline
Time scaling to compute $\ev{Z_i}_p$ & Tree & Non-tree  \\
\hline
Tensor-network contraction & $O(\exp(2 p))$   & $O(\exp(\exp((p)))$          \\
QAOA (all-to-all HW) & $O(p)$    & $O(p)$  \\
QAOA (planar HW) & $O(\exp( p)^{1/2})$ & $O(\exp( p)^{1/2})$  \\
\hline\hline
Set $p=\log(N)$ & Tree & Non-tree  \\
\hline
Tensor-network contraction & $O(N^2)$ & $O(\exp(N))$ \\
QAOA (all-to-all HW) & $O(\log(N))$ & $O(\log(N))$ \\
QAOA (planar HW)& $O(N^{1/2})$ & $O(N^{1/2})$ \\
\hline\hline
\end{tabular}
\end{center}
\caption{\textbf{Run-time scaling of computing the local expectation value $\ev{Z_i}_p$ as a function of the QAOA depth $p$.} The QAOA complexities are for computing $\ev{Z_i}_p$ up to additive precision, while tensor-network contraction is used as an exact classical method. }
\label{tab:cost}
\end{table}

\subsection{Implementation on IQM hardware} \label{subsec:implementation}

We implement the quantum-enhanced greedy algorithm on a 20-qubit superconducting quantum computer IQM Garnet~\cite{abdurakhimov2024}, with the qubit layout as in Fig.~\ref{fig:sketch}{(C)}. The expectation values $\ev{Z_i}_p$ are estimated on the quantum computer by repeatedly measuring the central qubit $i$ in state $|\bm{\gamma}^{\star}, \bm{\beta}^{\star}\rangle$ restricted to the light-cone subgraph. The states $|\bm{\gamma}^{\star}, \bm{\beta}^{\star}\rangle$ are prepared according to Eq.~\eqref{eq:qaoa_ansatz}, when the problem Hamiltonian encodes a specific subgraph. Namely, the $Z_i Z_j$ and $Z_i$ terms of Eq.~\eqref{eq:MIS_z} are defined by the respective edges and nodes of a given subgraph. Note that only gates belonging to the light-cone of the central qubit need to be implemented. This means that, labeling QAOA layers as $k =\{0,1,2,...,p-1\}$, the $k$-th QAOA layer will only include gates that are $p-k$ entangling gates away from the central qubit (see also Fig. 1 in Ref.~\cite{Wybo2025}). Finally, before execution, the ansätze $|\bm{\gamma}^{\star}, \bm{\beta}^{\star}\rangle$ are transpiled to IQM Garnet's native single qubit rotations and entangling CZ gates routed to its square-grid connectivity. These light-cone subgraph induced ansätze have up to $3 \cdot 2^p - 2$ qubits.

Fig.~\ref{fig:z_exp} shows $\ev{Z_i}_p$ of all possible subgraphs when the number of QAOA layers is $p=2$. These subgraphs correspond to all possible distance-2 neighbourhood light cones of a given node that can appear at any step of the algorithm. Such subgraphs consists of at most 10 nodes. Using the precomputed tree angles, we prepare and measure the quantum state of each subgraph a total of $20 000$ times. We employ Pauli twirling~\cite{wallman2016,hashim2020} to convert problematic coherent noise to more predictable depolarizing noise. Concretely, we create $20$ Pauli twirled versions from each circuit. These are executed $1000$ times each and the results are combined to get the total of $20 000$ measurement outcomes. We estimate the confidence intervals by showing standard deviation from bootstrapping. In bootstrapping we classically resample the counts from quantum circuit measurements $1000$ times to calculate the mean and standard deviation.

Fig.~\ref{fig:result} reveals that as long as the hierarchy between the expectation values remains sufficiently unaltered by noise, the quantum-enhanced greedy algorithm can make near-optimal choices and finds solutions comparable in quality to those from noiseless tensor network simulations for $p=2$. 

We also ran the $p=3$ algorithm (see Fig.~\ref{fig:result}), however in this case few subgraphs have 22 or 21 nodes which is more than the 20 qubits available on the IQM Garnet device. For nodes associated with these large subgraphs (one with 22 and three with 21 nodes), we assign noiseless expectation values in the quantum-enhanced greedy algorithm. Such nodes are rarely selected for inclusion in the independent set, and consequently the omission of device noise in these rare cases has no observable impact on the outcome of the algorithm. We find that the algorithm using \emph{noisy} expectation values from the device at depth $p=3$ performs similarly to the $p=2$ case. We discuss this feature, and the effects of noise more broadly in the next section.

\begin{figure*}[t]
  \centering
  \includegraphics[height=1.76in]{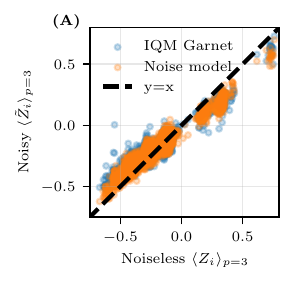}
  \includegraphics[height=1.76in]{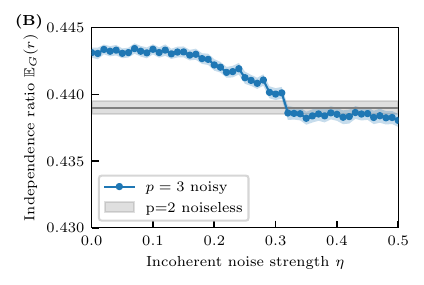}
  \includegraphics[height=1.76in]{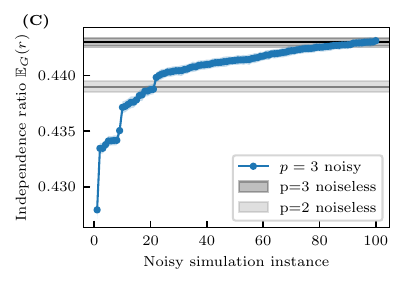}
  \caption{\textbf{(A) Correlation plot between the noisy and ideal expectation values} for $p=3$. The device measurements are shown in blue. The modeled values are shown in orange and are obtained via fitting Eq.~\eqref{eq:error-model} to the device measurements. The fit parameters are $\eta=0.03$, $\alpha=-0.05$ and $\sigma=0.04$. \textbf{(B) The performance of the quantum-enhanced greedy algorithm under shrinking noise} for $p=3$ and $N=1000$ as a function of $\eta$. The noise is modeled according to Eq.~\eqref{eq:shrinking}. \textbf{(C) The performance of the quantum-enhanced greedy algorithm under realistic noise} for $p=3$ and $N=1000$. The noise is modeled according to Eq.~\eqref{eq:error-model} with the same fit parameters as in (A). We consider 100 different random noise realizations of this model. We consider the same problem set of 200 graphs of size $N=1000$ in each of those. We have sorted the outcomes according to the achieved performances. }
  \label{fig:noise}
\end{figure*}

\subsection{Noise modeling} \label{subsec:noise}


To interpret the performance of the quantum-enhanced greedy algorithm on the IQM Garnet device, we introduce a simple phenomenological model for the measured local expectation values. Rather than attempting a full device-level noise simulation, we model how hardware noise distorts the ideal QAOA light-cone expectation values that drive the greedy decisions.

Specifically, we assume that the expectation value of the single-qubit observable $\langle Z_i\rangle$ obtained from the device can be modeled as a shrunken and biased version of the ideal value, plus residual fluctuations,
\begin{equation}
\label{eq:error-model}
    \langle\tilde{Z}_{i}\rangle
    =
    (1-\eta)^{|\tilde V_i|}\,\langle Z_i\rangle
    + 
    \alpha
    +
    \xi_{i}.
\end{equation}
Here, $|\tilde V_i|$ denotes the number of qubits in the light-cone subgraph used to evaluate $\langle Z_i\rangle$ at depth $p$ and $\xi_{i} \sim \mathcal{N}(0,\sigma)$ are the residual fluctuations. The first term models a subgraph-size dependent shrinking with rate $\eta$, motivated by the fact that the total number of imperfect operations (and hence the accumulated decoherence) increases with the size of the subgraph. Such shrinking of $\langle Z_i\rangle$ can be for instance due to depolarizing and dephasing noise. The offset $\alpha$ captures an effective, subgraph-size independent bias in the measured $\langle Z_i\rangle$ caused by e.g., amplitude damping or readout (assignment) errors~\cite{Funcke2022}.
Finally, the additive Gaussian term $\xi_{i} \sim \mathcal{N}(0,\sigma)$ models residual fluctuations, comprising shot noise as well as additional unmodeled error sources.

We fit the parameters of the error model in Eq.~\eqref{eq:error-model} to the hardware-measured expectation values obtained at $p=3$ (see Sec.~\ref{subsec:implementation}). Figure~\ref{fig:noise}(A) shows that the fitted model reproduces the dominant structure of the experimental data. The fitted deterministic parameters, $\eta=0.03$ and $\alpha=-0.05$, are roughly compatible with the gate fidelities of the device when taking into account the structure of the circuit decomposed into native gates (median CZ gate fidelity $0.997$). The inferred standard deviation of the residual term, $\sigma=0.04$, is substantially larger than what would be predicted from finite-shot statistics given our measurement budget ($1/\sqrt{20000}\approx 0.007\ll\sigma$). This suggests that the residual fluctuations are dominated by other effects not considered here.

We now investigate the impact of noise on the performance of the quantum-enhanced greedy algorithm. We first model the device noise by applying a multiplicative shrinking of the ideal expectation values obtained from tensor-network simulations at $p=3$, where the shrinking factor depends explicitly on the size of the corresponding light-cone subgraph. In other words, we only keep the first term in Eq.~\eqref{eq:error-model}, i.e.
\begin{equation} \label{eq:shrinking}
  \langle\tilde{Z}_i\rangle=(1-\eta)^{|\tilde V_i|}\,\langle Z_i\rangle
\end{equation}
 and vary $\eta$. This procedure captures the systematic shrinking of expectation values induced by for instance dephasing or depolarizing noise. As shown in Fig.~\ref{fig:noise}{(B)}, for fixed graph size $N=1000$, the performance of the quantum-enhanced greedy algorithm remains remarkably stable under this noise model and even seems to reach a stable region around the $p=2$ noiseless performance. However, this region is only reached for values of $\eta$ that are much larger than fitted value.

We therefore conclude that the dominant source of performance degradation observed in the device data cannot be attributed to subgraph-size dependent uniform shrinking alone, but is instead due to random distortions arising from shot noise and other stochastic error sources. To assess this effect, we additionally implemented the random noise model shown in Fig.~\ref{fig:noise}{(A)} parameterized by the fitted noise parameters $\eta$ and the fitted $\sigma$. Note that the uniform bias $\alpha$ does not alter the algorithm performance. For this model, we generated $100$ independent noise realizations characterized by $100$ distinct random seeds. Each realization consists of a sequence of random offsets added to the expectation values corresponding to the different subgraphs. So, the length of each sequence is the number of different subgraphs listed in Table~\ref{tab:counts}). For each such realization, we executed the greedy algorithm on the ensemble of 200 random graphs of fixed size $N=1000$. In Fig.~\ref{fig:noise}{(C)}, we indeed observe a strong sensitivity of the algorithmic performance to the specific noise realization. This is consistent with the performance obtained using the experimentally measured noisy expectation values, which display a similarly large spread (see Fig.~\ref{fig:noise}{(A)}).
Despite the large fluctuations, the observed performance shown in Fig.~\ref{fig:result} is captured by this model. This indicates that random fluctuations, in addition to subgraph-size dependent shrinking, are the dominant factors explaining the observed $p=3$ hardware performance.

\section{Discussion and comparison to classical algorithms} \label{sec:discussion}

We benchmarked the quantum-enhanced greedy algorithm against two classical baselines: (i) the state-of-the-art (SOTA) linear-prioritized search algorithm of Ref.~\cite{Marino2020} and (ii) the standard minimal-degree greedy heuristic (see Fig.~\ref{fig:result}). For the SOTA comparison we generated $10^5$ random $3$-regular graph instances for each system size $N$. Our results show that the greedy method combined with QAOA depth $p=4$ achieves a statistically significant improvement over the SOTA algorithm for all considered sizes up to $N\leq 2000$. This demonstrates that even a small number of QAOA layers, when combined with a lightweight classical reduction strategy, can already provide a significant performance boost over classical heuristics. Due to the monotonically improving performance of vanilla QAOA with $p$, further gains are expected from the quantum-enhanced algorithm as deeper QAOA circuits become accessible. While our finite-size simulations do not yet exceed the asymptotic value $r_{\infty}$ reported for the SOTA algorithm, it remains an interesting open question whether larger-size instances or higher QAOA depths could ultimately surpass this classical baseline.


Simulated Annealing (SA) provides an additional comparison point~\cite{Kirkpatrick1983}. Although SA is a heuristic with no performance guarantees, it is known to find comparatively large independent sets on regular graphs~\cite{Angelini2019}.  Its advantage comes from the ability to occasionally accept worse moves, allowing it to escape the local minima that trap greedy algorithms. In line with these prior observations, SA also tend to produce independent sets that exceed the asymptotic SOTA value $r_{\infty}$, but this improvement is purely heuristic (SA also does not exploit the structural locality of regular graphs in a principled way) and comes at the cost of increased `larger than linear' run time according to Ref.~\cite{Angelini2019}. In contrast, the quantum greedy algorithm obtains its guidance from QAOA expectation values, which encode quantum correlations arising from the underlying $3$-regular structure.

\section{Conclusion} \label{sec:concl}
In this work we investigated a quantum-enhanced greedy algorithm for finding large independent sets on 3-regular graphs, combining expectation values from QAOA with a greedy classical selection rule. By exploiting the local structure of the problem and reusing subgraph evaluations, we achieved average-case performance on random 3-regular graphs at shallow QAOA depths $p\approx 4$ that is competitive with the SOTA algorithm. In contrast, standard QAOA alone, does not even outperform simple greedy heuristics at these depths, thus highlighting the benefits of the hybrid algorithm.

We demonstrated the practical feasibility of this approach with experimental runs on quantum hardware, confirming that expectation values extracted from real devices can meaningfully improve upon the classical greedy decisions. While classical simulation is still possible at low depth using tensor-network methods, increasing the QAOA depth rapidly makes classical evaluation of expectation values infeasible due to exponential contraction cost. We have thus identified a regime of a potential quantum advantage, where quantum processors will become essential for carrying out the subgraph evaluations efficiently.

Furthermore, we emphasize a practical benefit of our scheme: the quantum greedy algorithm can be readily integrated into existing classical MIS solvers~\cite{Schuetz2025}. The quantum module supplies information about variable correlations at each iteration, while the classical reduction step ensures that the problem shrinks rapidly by enforcing constraints. This modularity could make the method compatible with powerful classical heuristics. 

Beyond MIS, we expect that this class of \emph{non-variational} quantum-enhanced approaches naturally extends to other combinatorial optimization problems defined on sparse graphs, where locality and a bounded-degree structure play a central role. Representative examples include MaxCut and related cut problems, minimum vertex cover, dominating set, and certain constraint satisfaction problems. In all these cases, relevant objective functions can be decomposed into local terms whose expectation values are accessible with shallow quantum circuits and can be incorporated into classical greedy or message-passing-style heuristics.

Taken together, our results indicate that modest quantum resources provide useful guidance for hybrid strategies of this type, suggesting a promising path toward stronger hybrid quantum-classical heuristics for MIS and beyond.

\section{Acknowledgements}

We thank Alessio Calzona, Jiri Guth Jarkovsky, Jalil Khatibi Moqadam, Miha Papič, Thomas Cope and Amin Hosseinkhani for helpful discussions and valuable feedback. We also acknowledge our colleagues at IQM for their support and for providing a collaborative research environment. This project is supported by the Federal Ministry for Economic Affairs and Climate Action on the basis of a decision by the German Bundestag through the project Quantum-enabling Services and Tools for Industrial Applications (QuaST). QuaST aims to facilitate the access to quantum-based solutions for optimization problems.

\bibliography{biblio}

\begin{thebibliography}{74}%
\makeatletter
\providecommand \@ifxundefined [1]{%
 \@ifx{#1\undefined}
}%
\providecommand \@ifnum [1]{%
 \ifnum #1\expandafter \@firstoftwo
 \else \expandafter \@secondoftwo
 \fi
}%
\providecommand \@ifx [1]{%
 \ifx #1\expandafter \@firstoftwo
 \else \expandafter \@secondoftwo
 \fi
}%
\providecommand \natexlab [1]{#1}%
\providecommand \enquote  [1]{``#1''}%
\providecommand \bibnamefont  [1]{#1}%
\providecommand \bibfnamefont [1]{#1}%
\providecommand \citenamefont [1]{#1}%
\providecommand \href@noop [0]{\@secondoftwo}%
\providecommand \href [0]{\begingroup \@sanitize@url \@href}%
\providecommand \@href[1]{\@@startlink{#1}\@@href}%
\providecommand \@@href[1]{\endgroup#1\@@endlink}%
\providecommand \@sanitize@url [0]{\catcode `\\12\catcode `\$12\catcode
  `\&12\catcode `\#12\catcode `\^12\catcode `\_12\catcode `\%12\relax}%
\providecommand \@@startlink[1]{}%
\providecommand \@@endlink[0]{}%
\providecommand \url  [0]{\begingroup\@sanitize@url \@url }%
\providecommand \@url [1]{\endgroup\@href {#1}{\urlprefix }}%
\providecommand \urlprefix  [0]{URL }%
\providecommand \Eprint [0]{\href }%
\providecommand \doibase [0]{http://dx.doi.org/}%
\providecommand \selectlanguage [0]{\@gobble}%
\providecommand \bibinfo  [0]{\@secondoftwo}%
\providecommand \bibfield  [0]{\@secondoftwo}%
\providecommand \translation [1]{[#1]}%
\providecommand \BibitemOpen [0]{}%
\providecommand \bibitemStop [0]{}%
\providecommand \bibitemNoStop [0]{.\EOS\space}%
\providecommand \EOS [0]{\spacefactor3000\relax}%
\providecommand \BibitemShut  [1]{\csname bibitem#1\endcsname}%
\let\auto@bib@innerbib\@empty
\bibitem [{\citenamefont {Abbas}\ \emph {et~al.}(2024)\citenamefont {Abbas},
  \citenamefont {Ambainis}, \citenamefont {Augustino}, \citenamefont
  {B\"{a}rtschi}, \citenamefont {Buhrman}, \citenamefont {Coffrin},
  \citenamefont {Cortiana}, \citenamefont {Dunjko}, \citenamefont {Egger},
  \citenamefont {Elmegreen}, \citenamefont {Franco}, \citenamefont {Fratini},
  \citenamefont {Fuller}, \citenamefont {Gacon}, \citenamefont {Gonciulea},
  \citenamefont {Gribling}, \citenamefont {Gupta}, \citenamefont {Hadfield},
  \citenamefont {Heese}, \citenamefont {Kircher}, \citenamefont {Kleinert},
  \citenamefont {Koch}, \citenamefont {Korpas}, \citenamefont {Lenk},
  \citenamefont {Marecek}, \citenamefont {Markov}, \citenamefont {Mazzola},
  \citenamefont {Mensa}, \citenamefont {Mohseni}, \citenamefont {Nannicini},
  \citenamefont {O’Meara}, \citenamefont {Tapia}, \citenamefont {Pokutta},
  \citenamefont {Proissl}, \citenamefont {Rebentrost}, \citenamefont {Sahin},
  \citenamefont {Symons}, \citenamefont {Tornow}, \citenamefont {Valls},
  \citenamefont {Woerner}, \citenamefont {Wolf-Bauwens}, \citenamefont {Yard},
  \citenamefont {Yarkoni}, \citenamefont {Zechiel}, \citenamefont {Zhuk},\ and\
  \citenamefont {Zoufal}}]{Abbas2024}%
  \BibitemOpen
  \bibfield  {author} {\bibinfo {author} {\bibfnamefont {A.}~\bibnamefont
  {Abbas}}, \bibinfo {author} {\bibfnamefont {A.}~\bibnamefont {Ambainis}},
  \bibinfo {author} {\bibfnamefont {B.}~\bibnamefont {Augustino}}, \bibinfo
  {author} {\bibfnamefont {A.}~\bibnamefont {B\"{a}rtschi}}, \bibinfo {author}
  {\bibfnamefont {H.}~\bibnamefont {Buhrman}}, \bibinfo {author} {\bibfnamefont
  {C.}~\bibnamefont {Coffrin}}, \bibinfo {author} {\bibfnamefont
  {G.}~\bibnamefont {Cortiana}}, \bibinfo {author} {\bibfnamefont
  {V.}~\bibnamefont {Dunjko}}, \bibinfo {author} {\bibfnamefont {D.~J.}\
  \bibnamefont {Egger}}, \bibinfo {author} {\bibfnamefont {B.~G.}\ \bibnamefont
  {Elmegreen}}, \bibinfo {author} {\bibfnamefont {N.}~\bibnamefont {Franco}},
  \bibinfo {author} {\bibfnamefont {F.}~\bibnamefont {Fratini}}, \bibinfo
  {author} {\bibfnamefont {B.}~\bibnamefont {Fuller}}, \bibinfo {author}
  {\bibfnamefont {J.}~\bibnamefont {Gacon}}, \bibinfo {author} {\bibfnamefont
  {C.}~\bibnamefont {Gonciulea}}, \bibinfo {author} {\bibfnamefont
  {S.}~\bibnamefont {Gribling}}, \bibinfo {author} {\bibfnamefont
  {S.}~\bibnamefont {Gupta}}, \bibinfo {author} {\bibfnamefont
  {S.}~\bibnamefont {Hadfield}}, \bibinfo {author} {\bibfnamefont
  {R.}~\bibnamefont {Heese}}, \bibinfo {author} {\bibfnamefont
  {G.}~\bibnamefont {Kircher}}, \bibinfo {author} {\bibfnamefont
  {T.}~\bibnamefont {Kleinert}}, \bibinfo {author} {\bibfnamefont
  {T.}~\bibnamefont {Koch}}, \bibinfo {author} {\bibfnamefont {G.}~\bibnamefont
  {Korpas}}, \bibinfo {author} {\bibfnamefont {S.}~\bibnamefont {Lenk}},
  \bibinfo {author} {\bibfnamefont {J.}~\bibnamefont {Marecek}}, \bibinfo
  {author} {\bibfnamefont {V.}~\bibnamefont {Markov}}, \bibinfo {author}
  {\bibfnamefont {G.}~\bibnamefont {Mazzola}}, \bibinfo {author} {\bibfnamefont
  {S.}~\bibnamefont {Mensa}}, \bibinfo {author} {\bibfnamefont
  {N.}~\bibnamefont {Mohseni}}, \bibinfo {author} {\bibfnamefont
  {G.}~\bibnamefont {Nannicini}}, \bibinfo {author} {\bibfnamefont
  {C.}~\bibnamefont {O’Meara}}, \bibinfo {author} {\bibfnamefont {E.~P.}\
  \bibnamefont {Tapia}}, \bibinfo {author} {\bibfnamefont {S.}~\bibnamefont
  {Pokutta}}, \bibinfo {author} {\bibfnamefont {M.}~\bibnamefont {Proissl}},
  \bibinfo {author} {\bibfnamefont {P.}~\bibnamefont {Rebentrost}}, \bibinfo
  {author} {\bibfnamefont {E.}~\bibnamefont {Sahin}}, \bibinfo {author}
  {\bibfnamefont {B.~C.~B.}\ \bibnamefont {Symons}}, \bibinfo {author}
  {\bibfnamefont {S.}~\bibnamefont {Tornow}}, \bibinfo {author} {\bibfnamefont
  {V.}~\bibnamefont {Valls}}, \bibinfo {author} {\bibfnamefont
  {S.}~\bibnamefont {Woerner}}, \bibinfo {author} {\bibfnamefont {M.~L.}\
  \bibnamefont {Wolf-Bauwens}}, \bibinfo {author} {\bibfnamefont
  {J.}~\bibnamefont {Yard}}, \bibinfo {author} {\bibfnamefont {S.}~\bibnamefont
  {Yarkoni}}, \bibinfo {author} {\bibfnamefont {D.}~\bibnamefont {Zechiel}},
  \bibinfo {author} {\bibfnamefont {S.}~\bibnamefont {Zhuk}}, \ and\ \bibinfo
  {author} {\bibfnamefont {C.}~\bibnamefont {Zoufal}},\ }\href {\doibase
  10.1038/s42254-024-00770-9} {\bibfield  {journal} {\bibinfo  {journal}
  {Nature Reviews Physics}\ }\textbf {\bibinfo {volume} {6}},\ \bibinfo {pages}
  {718–735} (\bibinfo {year} {2024})}\BibitemShut {NoStop}%
\bibitem [{\citenamefont {Preskill}(2018)}]{Preskill2018}%
  \BibitemOpen
  \bibfield  {author} {\bibinfo {author} {\bibfnamefont {J.}~\bibnamefont
  {Preskill}},\ }\href {\doibase 10.22331/q-2018-08-06-79} {\bibfield
  {journal} {\bibinfo  {journal} {Quantum 2, 79 (2018)}\ }\textbf {\bibinfo
  {volume} {2}},\ \bibinfo {pages} {79} (\bibinfo {year} {2018})},\ \Eprint
  {http://arxiv.org/abs/1801.00862} {arXiv:1801.00862 [quant-ph]} \BibitemShut
  {NoStop}%
\bibitem [{\citenamefont {Farhi}\ \emph {et~al.}(2014)\citenamefont {Farhi},
  \citenamefont {Goldstone},\ and\ \citenamefont {Gutmann}}]{Farhi2014}%
  \BibitemOpen
  \bibfield  {author} {\bibinfo {author} {\bibfnamefont {E.}~\bibnamefont
  {Farhi}}, \bibinfo {author} {\bibfnamefont {J.}~\bibnamefont {Goldstone}}, \
  and\ \bibinfo {author} {\bibfnamefont {S.}~\bibnamefont {Gutmann}},\ }\href
  {\doibase 10.48550/ARXIV.1411.4028} {\  (\bibinfo {year} {2014}),\
  10.48550/ARXIV.1411.4028},\ \Eprint {http://arxiv.org/abs/1411.4028}
  {arXiv:1411.4028 [quant-ph]} \BibitemShut {NoStop}%
\bibitem [{\citenamefont {Blekos}\ \emph {et~al.}(2024)\citenamefont {Blekos},
  \citenamefont {Brand}, \citenamefont {Ceschini}, \citenamefont {Chou},
  \citenamefont {Li}, \citenamefont {Pandya},\ and\ \citenamefont
  {Summer}}]{Blekos2024}%
  \BibitemOpen
  \bibfield  {author} {\bibinfo {author} {\bibfnamefont {K.}~\bibnamefont
  {Blekos}}, \bibinfo {author} {\bibfnamefont {D.}~\bibnamefont {Brand}},
  \bibinfo {author} {\bibfnamefont {A.}~\bibnamefont {Ceschini}}, \bibinfo
  {author} {\bibfnamefont {C.-H.}\ \bibnamefont {Chou}}, \bibinfo {author}
  {\bibfnamefont {R.-H.}\ \bibnamefont {Li}}, \bibinfo {author} {\bibfnamefont
  {K.}~\bibnamefont {Pandya}}, \ and\ \bibinfo {author} {\bibfnamefont
  {A.}~\bibnamefont {Summer}},\ }\href {\doibase 10.1016/j.physrep.2024.03.002}
  {\bibfield  {journal} {\bibinfo  {journal} {Physics Reports}\ }\textbf
  {\bibinfo {volume} {1068}},\ \bibinfo {pages} {1–66} (\bibinfo {year}
  {2024})}\BibitemShut {NoStop}%
\bibitem [{\citenamefont {Brandao}\ \emph {et~al.}(2018)\citenamefont
  {Brandao}, \citenamefont {Broughton}, \citenamefont {Farhi}, \citenamefont
  {Gutmann},\ and\ \citenamefont {Neven}}]{Brandao2018}%
  \BibitemOpen
  \bibfield  {author} {\bibinfo {author} {\bibfnamefont {F.~G. S.~L.}\
  \bibnamefont {Brandao}}, \bibinfo {author} {\bibfnamefont {M.}~\bibnamefont
  {Broughton}}, \bibinfo {author} {\bibfnamefont {E.}~\bibnamefont {Farhi}},
  \bibinfo {author} {\bibfnamefont {S.}~\bibnamefont {Gutmann}}, \ and\
  \bibinfo {author} {\bibfnamefont {H.}~\bibnamefont {Neven}},\ }\href
  {\doibase 10.48550/ARXIV.1812.04170} {\  (\bibinfo {year} {2018}),\
  10.48550/ARXIV.1812.04170},\ \Eprint {http://arxiv.org/abs/1812.04170}
  {arXiv:1812.04170 [quant-ph]} \BibitemShut {NoStop}%
\bibitem [{\citenamefont {Wang}\ \emph {et~al.}(2018)\citenamefont {Wang},
  \citenamefont {Hadfield}, \citenamefont {Jiang},\ and\ \citenamefont
  {Rieffel}}]{Wang2018}%
  \BibitemOpen
  \bibfield  {author} {\bibinfo {author} {\bibfnamefont {Z.}~\bibnamefont
  {Wang}}, \bibinfo {author} {\bibfnamefont {S.}~\bibnamefont {Hadfield}},
  \bibinfo {author} {\bibfnamefont {Z.}~\bibnamefont {Jiang}}, \ and\ \bibinfo
  {author} {\bibfnamefont {E.~G.}\ \bibnamefont {Rieffel}},\ }\href {\doibase
  10.1103/PhysRevA.97.022304} {\bibfield  {journal} {\bibinfo  {journal} {Phys.
  Rev. A}\ }\textbf {\bibinfo {volume} {97}},\ \bibinfo {pages} {022304}
  (\bibinfo {year} {2018})}\BibitemShut {NoStop}%
\bibitem [{\citenamefont {Basso}\ \emph {et~al.}(2021)\citenamefont {Basso},
  \citenamefont {Farhi}, \citenamefont {Marwaha}, \citenamefont {Villalonga},\
  and\ \citenamefont {Zhou}}]{Basso2021}%
  \BibitemOpen
  \bibfield  {author} {\bibinfo {author} {\bibfnamefont {J.}~\bibnamefont
  {Basso}}, \bibinfo {author} {\bibfnamefont {E.}~\bibnamefont {Farhi}},
  \bibinfo {author} {\bibfnamefont {K.}~\bibnamefont {Marwaha}}, \bibinfo
  {author} {\bibfnamefont {B.}~\bibnamefont {Villalonga}}, \ and\ \bibinfo
  {author} {\bibfnamefont {L.}~\bibnamefont {Zhou}},\ }\href {\doibase
  10.4230/LIPICS.TQC.2022.7} {\bibfield  {journal} {\bibinfo  {journal} {In
  Proceedings of the 17th Conference on the Theory of Quantum Computation,
  Communication and Cryptography (TQC '22), 7:1--7:21, (2022)}\ } (\bibinfo
  {year} {2021}),\ 10.4230/LIPICS.TQC.2022.7},\ \Eprint
  {http://arxiv.org/abs/2110.14206} {arXiv:2110.14206 [quant-ph]} \BibitemShut
  {NoStop}%
\bibitem [{\citenamefont {Farhi}\ \emph
  {et~al.}(2020{\natexlab{a}})\citenamefont {Farhi}, \citenamefont {Gamarnik},\
  and\ \citenamefont {Gutmann}}]{Farhi2020}%
  \BibitemOpen
  \bibfield  {author} {\bibinfo {author} {\bibfnamefont {E.}~\bibnamefont
  {Farhi}}, \bibinfo {author} {\bibfnamefont {D.}~\bibnamefont {Gamarnik}}, \
  and\ \bibinfo {author} {\bibfnamefont {S.}~\bibnamefont {Gutmann}},\ }\href
  {\doibase 10.48550/ARXIV.2005.08747} {\  (\bibinfo {year}
  {2020}{\natexlab{a}}),\ 10.48550/ARXIV.2005.08747},\ \Eprint
  {http://arxiv.org/abs/2005.08747} {arXiv:2005.08747 [quant-ph]} \BibitemShut
  {NoStop}%
\bibitem [{\citenamefont {Farhi}\ \emph
  {et~al.}(2020{\natexlab{b}})\citenamefont {Farhi}, \citenamefont {Gamarnik},\
  and\ \citenamefont {Gutmann}}]{Farhi2020a}%
  \BibitemOpen
  \bibfield  {author} {\bibinfo {author} {\bibfnamefont {E.}~\bibnamefont
  {Farhi}}, \bibinfo {author} {\bibfnamefont {D.}~\bibnamefont {Gamarnik}}, \
  and\ \bibinfo {author} {\bibfnamefont {S.}~\bibnamefont {Gutmann}},\ }\href
  {\doibase 10.48550/ARXIV.2004.09002} {\  (\bibinfo {year}
  {2020}{\natexlab{b}}),\ 10.48550/ARXIV.2004.09002},\ \Eprint
  {http://arxiv.org/abs/2004.09002} {arXiv:2004.09002 [quant-ph]} \BibitemShut
  {NoStop}%
\bibitem [{\citenamefont {Farhi}\ \emph {et~al.}(2022)\citenamefont {Farhi},
  \citenamefont {Goldstone}, \citenamefont {Gutmann},\ and\ \citenamefont
  {Zhou}}]{Farhi2022}%
  \BibitemOpen
  \bibfield  {author} {\bibinfo {author} {\bibfnamefont {E.}~\bibnamefont
  {Farhi}}, \bibinfo {author} {\bibfnamefont {J.}~\bibnamefont {Goldstone}},
  \bibinfo {author} {\bibfnamefont {S.}~\bibnamefont {Gutmann}}, \ and\
  \bibinfo {author} {\bibfnamefont {L.}~\bibnamefont {Zhou}},\ }\href {\doibase
  10.22331/q-2022-07-07-759} {\bibfield  {journal} {\bibinfo  {journal}
  {Quantum}\ }\textbf {\bibinfo {volume} {6}},\ \bibinfo {pages} {759}
  (\bibinfo {year} {2022})}\BibitemShut {NoStop}%
\bibitem [{\citenamefont {Zhou}\ \emph {et~al.}(2018)\citenamefont {Zhou},
  \citenamefont {Wang}, \citenamefont {Choi}, \citenamefont {Pichler},\ and\
  \citenamefont {Lukin}}]{Zhou2018}%
  \BibitemOpen
  \bibfield  {author} {\bibinfo {author} {\bibfnamefont {L.}~\bibnamefont
  {Zhou}}, \bibinfo {author} {\bibfnamefont {S.-T.}\ \bibnamefont {Wang}},
  \bibinfo {author} {\bibfnamefont {S.}~\bibnamefont {Choi}}, \bibinfo {author}
  {\bibfnamefont {H.}~\bibnamefont {Pichler}}, \ and\ \bibinfo {author}
  {\bibfnamefont {M.~D.}\ \bibnamefont {Lukin}},\ }\href {\doibase
  10.1103/physrevx.10.021067} {\bibfield  {journal} {\bibinfo  {journal} {Phys.
  Rev. X 10, 021067 (2020)}\ }\textbf {\bibinfo {volume} {10}},\ \bibinfo
  {pages} {021067} (\bibinfo {year} {2018})},\ \Eprint
  {http://arxiv.org/abs/1812.01041} {arXiv:1812.01041 [quant-ph]} \BibitemShut
  {NoStop}%
\bibitem [{\citenamefont {Pagano}\ and\ \citenamefont {{\em et
  al.}}(2020)}]{Pagano2020}%
  \BibitemOpen
  \bibfield  {author} {\bibinfo {author} {\bibfnamefont {G.}~\bibnamefont
  {Pagano}}\ and\ \bibinfo {author} {\bibnamefont {{\em et al.}}},\ }\href
  {https://pmc.ncbi.nlm.nih.gov/articles/PMC7568299/} {\bibfield  {journal}
  {\bibinfo  {journal} {Proc. Natl. Acad. Sci. U.S.A.}\ } (\bibinfo {year}
  {2020})},\ \bibinfo {note} {trapped-ion experimental implementation of QAOA
  on long-range Ising models}\BibitemShut {NoStop}%
\bibitem [{\citenamefont {Harrigan}\ \emph {et~al.}(2021)\citenamefont
  {Harrigan}, \citenamefont {Sung},\ and\ \citenamefont {\textit{et
  al.}}}]{Harrigan2021}%
  \BibitemOpen
  \bibfield  {author} {\bibinfo {author} {\bibfnamefont {M.~P.}\ \bibnamefont
  {Harrigan}}, \bibinfo {author} {\bibfnamefont {K.~J.}\ \bibnamefont {Sung}},
  \ and\ \bibinfo {author} {\bibfnamefont {M.~N.}\ \bibnamefont {\textit{et
  al.}}},\ }\href@noop {} {\bibfield  {journal} {\bibinfo  {journal} {Nature
  Phys.}\ }\textbf {\bibinfo {volume} {17}},\ \bibinfo {pages} {332} (\bibinfo
  {year} {2021})}\BibitemShut {NoStop}%
\bibitem [{\citenamefont {Ebadi}\ \emph {et~al.}(2022)\citenamefont {Ebadi},
  \citenamefont {Keesling}, \citenamefont {Cain}, \citenamefont {Wang},
  \citenamefont {Levine}, \citenamefont {Bluvstein}, \citenamefont {Semeghini},
  \citenamefont {Omran}, \citenamefont {Liu}, \citenamefont {Samajdar},
  \citenamefont {Luo}, \citenamefont {Nash}, \citenamefont {Gao}, \citenamefont
  {Barak}, \citenamefont {Farhi}, \citenamefont {Sachdev}, \citenamefont
  {Gemelke}, \citenamefont {Zhou}, \citenamefont {Choi}, \citenamefont
  {Pichler}, \citenamefont {Wang}, \citenamefont {Greiner}, \citenamefont
  {Vuletić},\ and\ \citenamefont {Lukin}}]{Ebadi2022}%
  \BibitemOpen
  \bibfield  {author} {\bibinfo {author} {\bibfnamefont {S.}~\bibnamefont
  {Ebadi}}, \bibinfo {author} {\bibfnamefont {A.}~\bibnamefont {Keesling}},
  \bibinfo {author} {\bibfnamefont {M.}~\bibnamefont {Cain}}, \bibinfo {author}
  {\bibfnamefont {T.~T.}\ \bibnamefont {Wang}}, \bibinfo {author}
  {\bibfnamefont {H.}~\bibnamefont {Levine}}, \bibinfo {author} {\bibfnamefont
  {D.}~\bibnamefont {Bluvstein}}, \bibinfo {author} {\bibfnamefont
  {G.}~\bibnamefont {Semeghini}}, \bibinfo {author} {\bibfnamefont
  {A.}~\bibnamefont {Omran}}, \bibinfo {author} {\bibfnamefont {J.-G.}\
  \bibnamefont {Liu}}, \bibinfo {author} {\bibfnamefont {R.}~\bibnamefont
  {Samajdar}}, \bibinfo {author} {\bibfnamefont {X.-Z.}\ \bibnamefont {Luo}},
  \bibinfo {author} {\bibfnamefont {B.}~\bibnamefont {Nash}}, \bibinfo {author}
  {\bibfnamefont {X.}~\bibnamefont {Gao}}, \bibinfo {author} {\bibfnamefont
  {B.}~\bibnamefont {Barak}}, \bibinfo {author} {\bibfnamefont
  {E.}~\bibnamefont {Farhi}}, \bibinfo {author} {\bibfnamefont
  {S.}~\bibnamefont {Sachdev}}, \bibinfo {author} {\bibfnamefont
  {N.}~\bibnamefont {Gemelke}}, \bibinfo {author} {\bibfnamefont
  {L.}~\bibnamefont {Zhou}}, \bibinfo {author} {\bibfnamefont {S.}~\bibnamefont
  {Choi}}, \bibinfo {author} {\bibfnamefont {H.}~\bibnamefont {Pichler}},
  \bibinfo {author} {\bibfnamefont {S.-T.}\ \bibnamefont {Wang}}, \bibinfo
  {author} {\bibfnamefont {M.}~\bibnamefont {Greiner}}, \bibinfo {author}
  {\bibfnamefont {V.}~\bibnamefont {Vuletić}}, \ and\ \bibinfo {author}
  {\bibfnamefont {M.~D.}\ \bibnamefont {Lukin}},\ }\href {\doibase
  10.1126/science.abo6587} {\bibfield  {journal} {\bibinfo  {journal}
  {Science}\ }\textbf {\bibinfo {volume} {376}},\ \bibinfo {pages}
  {1209–1215} (\bibinfo {year} {2022})}\BibitemShut {NoStop}%
\bibitem [{\citenamefont {Byun}\ \emph {et~al.}(2022)\citenamefont {Byun},
  \citenamefont {Kim},\ and\ \citenamefont {Ahn}}]{Byun2022}%
  \BibitemOpen
  \bibfield  {author} {\bibinfo {author} {\bibfnamefont {A.}~\bibnamefont
  {Byun}}, \bibinfo {author} {\bibfnamefont {M.}~\bibnamefont {Kim}}, \ and\
  \bibinfo {author} {\bibfnamefont {J.}~\bibnamefont {Ahn}},\ }\href {\doibase
  10.1103/PRXQuantum.3.030305} {\bibfield  {journal} {\bibinfo  {journal} {PRX
  Quantum}\ }\textbf {\bibinfo {volume} {3}},\ \bibinfo {pages} {030305}
  (\bibinfo {year} {2022})}\BibitemShut {NoStop}%
\bibitem [{\citenamefont {Kim}\ \emph {et~al.}(2022)\citenamefont {Kim},
  \citenamefont {Kim}, \citenamefont {Hwang}, \citenamefont {Moon},\ and\
  \citenamefont {Ahn}}]{Kim2022}%
  \BibitemOpen
  \bibfield  {author} {\bibinfo {author} {\bibfnamefont {M.}~\bibnamefont
  {Kim}}, \bibinfo {author} {\bibfnamefont {K.}~\bibnamefont {Kim}}, \bibinfo
  {author} {\bibfnamefont {J.}~\bibnamefont {Hwang}}, \bibinfo {author}
  {\bibfnamefont {E.-G.}\ \bibnamefont {Moon}}, \ and\ \bibinfo {author}
  {\bibfnamefont {J.}~\bibnamefont {Ahn}},\ }\href {\doibase
  10.1038/s41567-022-01629-5} {\bibfield  {journal} {\bibinfo  {journal}
  {Nature Physics}\ }\textbf {\bibinfo {volume} {18}},\ \bibinfo {pages}
  {755–759} (\bibinfo {year} {2022})}\BibitemShut {NoStop}%
\bibitem [{\citenamefont {Farhi}\ \emph {et~al.}(2025)\citenamefont {Farhi},
  \citenamefont {Gutmann}, \citenamefont {Ranard},\ and\ \citenamefont
  {Villalonga}}]{Farhi2025}%
  \BibitemOpen
  \bibfield  {author} {\bibinfo {author} {\bibfnamefont {E.}~\bibnamefont
  {Farhi}}, \bibinfo {author} {\bibfnamefont {S.}~\bibnamefont {Gutmann}},
  \bibinfo {author} {\bibfnamefont {D.}~\bibnamefont {Ranard}}, \ and\ \bibinfo
  {author} {\bibfnamefont {B.}~\bibnamefont {Villalonga}},\ }\href
  {https://arxiv.org/abs/2503.12789} {\enquote {\bibinfo {title} {Lower
  bounding the maxcut of high girth 3-regular graphs using the qaoa},}\ }
  (\bibinfo {year} {2025}),\ \Eprint {http://arxiv.org/abs/2503.12789}
  {arXiv:2503.12789 [quant-ph]} \BibitemShut {NoStop}%
\bibitem [{\citenamefont {Goemans}\ and\ \citenamefont
  {Williamson}(1995)}]{Goemans1995}%
  \BibitemOpen
  \bibfield  {author} {\bibinfo {author} {\bibfnamefont {M.~X.}\ \bibnamefont
  {Goemans}}\ and\ \bibinfo {author} {\bibfnamefont {D.~P.}\ \bibnamefont
  {Williamson}},\ }\href {\doibase 10.1145/227683.227684} {\bibfield  {journal}
  {\bibinfo  {journal} {Journal of the ACM}\ }\textbf {\bibinfo {volume}
  {42}},\ \bibinfo {pages} {1115–1145} (\bibinfo {year} {1995})}\BibitemShut
  {NoStop}%
\bibitem [{\citenamefont {Crooks}(2018)}]{Crooks2018}%
  \BibitemOpen
  \bibfield  {author} {\bibinfo {author} {\bibfnamefont {G.~E.}\ \bibnamefont
  {Crooks}},\ }\href {\doibase 10.48550/ARXIV.1811.08419} {\  (\bibinfo {year}
  {2018}),\ 10.48550/ARXIV.1811.08419},\ \Eprint
  {http://arxiv.org/abs/1811.08419} {arXiv:1811.08419 [quant-ph]} \BibitemShut
  {NoStop}%
\bibitem [{\citenamefont {Wurtz}\ and\ \citenamefont
  {Lykov}(2021)}]{Wurtz2021}%
  \BibitemOpen
  \bibfield  {author} {\bibinfo {author} {\bibfnamefont {J.}~\bibnamefont
  {Wurtz}}\ and\ \bibinfo {author} {\bibfnamefont {D.}~\bibnamefont {Lykov}},\
  }\href {\doibase 10.1103/PhysRevA.104.052419} {\bibfield  {journal} {\bibinfo
   {journal} {Phys. Rev. A}\ }\textbf {\bibinfo {volume} {104}},\ \bibinfo
  {pages} {052419} (\bibinfo {year} {2021})}\BibitemShut {NoStop}%
\bibitem [{\citenamefont {Chou}\ \emph {et~al.}(2021)\citenamefont {Chou},
  \citenamefont {Love}, \citenamefont {Sandhu},\ and\ \citenamefont
  {Shi}}]{Chou2021}%
  \BibitemOpen
  \bibfield  {author} {\bibinfo {author} {\bibfnamefont {C.-N.}\ \bibnamefont
  {Chou}}, \bibinfo {author} {\bibfnamefont {P.~J.}\ \bibnamefont {Love}},
  \bibinfo {author} {\bibfnamefont {J.~S.}\ \bibnamefont {Sandhu}}, \ and\
  \bibinfo {author} {\bibfnamefont {J.}~\bibnamefont {Shi}},\ }\href {\doibase
  10.48550/ARXIV.2108.06049} {\  (\bibinfo {year} {2021}),\
  10.48550/ARXIV.2108.06049},\ \Eprint {http://arxiv.org/abs/2108.06049}
  {arXiv:2108.06049 [quant-ph]} \BibitemShut {NoStop}%
\bibitem [{\citenamefont {Barak}\ and\ \citenamefont
  {Marwaha}(2021)}]{Barak2021}%
  \BibitemOpen
  \bibfield  {author} {\bibinfo {author} {\bibfnamefont {B.}~\bibnamefont
  {Barak}}\ and\ \bibinfo {author} {\bibfnamefont {K.}~\bibnamefont
  {Marwaha}},\ }\href {\doibase 10.4230/LIPICS.ITCS.2022.14} {\bibfield
  {journal} {\bibinfo  {journal} {13th Innovations in Theoretical Computer
  Science Conference (ITCS 2022); Article No. 14}\ } (\bibinfo {year} {2021}),\
  10.4230/LIPICS.ITCS.2022.14},\ \Eprint {http://arxiv.org/abs/2106.05900}
  {arXiv:2106.05900 [quant-ph]} \BibitemShut {NoStop}%
\bibitem [{\citenamefont {Anshu}\ and\ \citenamefont
  {Metger}(2022)}]{Anshu2022}%
  \BibitemOpen
  \bibfield  {author} {\bibinfo {author} {\bibfnamefont {A.}~\bibnamefont
  {Anshu}}\ and\ \bibinfo {author} {\bibfnamefont {T.}~\bibnamefont {Metger}},\
  }\href {\doibase 10.22331/q-2023-05-11-999} {\bibfield  {journal} {\bibinfo
  {journal} {Quantum 7, 999 (2023)}\ }\textbf {\bibinfo {volume} {7}},\
  \bibinfo {pages} {999} (\bibinfo {year} {2022})},\ \Eprint
  {http://arxiv.org/abs/2209.02715} {arXiv:2209.02715 [quant-ph]} \BibitemShut
  {NoStop}%
\bibitem [{\citenamefont {Chen}\ \emph {et~al.}(2023)\citenamefont {Chen},
  \citenamefont {Huang},\ and\ \citenamefont {Marwaha}}]{Chen2023}%
  \BibitemOpen
  \bibfield  {author} {\bibinfo {author} {\bibfnamefont {A.}~\bibnamefont
  {Chen}}, \bibinfo {author} {\bibfnamefont {N.}~\bibnamefont {Huang}}, \ and\
  \bibinfo {author} {\bibfnamefont {K.}~\bibnamefont {Marwaha}},\ }\href
  {\doibase 10.48550/ARXIV.2310.01563} {\  (\bibinfo {year} {2023}),\
  10.48550/ARXIV.2310.01563},\ \Eprint {http://arxiv.org/abs/2310.01563}
  {arXiv:2310.01563 [quant-ph]} \BibitemShut {NoStop}%
\bibitem [{\citenamefont {Franca}\ and\ \citenamefont
  {Garcia-Patron}(2020)}]{Franca2020}%
  \BibitemOpen
  \bibfield  {author} {\bibinfo {author} {\bibfnamefont {D.~S.}\ \bibnamefont
  {Franca}}\ and\ \bibinfo {author} {\bibfnamefont {R.}~\bibnamefont
  {Garcia-Patron}},\ }\href {\doibase 10.1038/s41567-021-01356-3} {\bibfield
  {journal} {\bibinfo  {journal} {Nature Physics}\ }\textbf {\bibinfo {volume}
  {17}},\ \bibinfo {pages} {1221} (\bibinfo {year} {2020})},\ \Eprint
  {http://arxiv.org/abs/2009.05532} {arXiv:2009.05532 [quant-ph]} \BibitemShut
  {NoStop}%
\bibitem [{\citenamefont {Wang}\ \emph {et~al.}(2021)\citenamefont {Wang},
  \citenamefont {Fontana}, \citenamefont {Cerezo}, \citenamefont {Sharma},
  \citenamefont {Sone}, \citenamefont {Cincio},\ and\ \citenamefont
  {Coles}}]{Wang2021}%
  \BibitemOpen
  \bibfield  {author} {\bibinfo {author} {\bibfnamefont {S.}~\bibnamefont
  {Wang}}, \bibinfo {author} {\bibfnamefont {E.}~\bibnamefont {Fontana}},
  \bibinfo {author} {\bibfnamefont {M.}~\bibnamefont {Cerezo}}, \bibinfo
  {author} {\bibfnamefont {K.}~\bibnamefont {Sharma}}, \bibinfo {author}
  {\bibfnamefont {A.}~\bibnamefont {Sone}}, \bibinfo {author} {\bibfnamefont
  {L.}~\bibnamefont {Cincio}}, \ and\ \bibinfo {author} {\bibfnamefont {P.~J.}\
  \bibnamefont {Coles}},\ }\href {\doibase 10.1038/s41467-021-27045-8}
  {\bibfield  {journal} {\bibinfo  {journal} {Nature Communications}\ }\textbf
  {\bibinfo {volume} {12}},\ \bibinfo {pages} {6961} (\bibinfo {year}
  {2021})}\BibitemShut {NoStop}%
\bibitem [{\citenamefont {De~Palma}\ \emph {et~al.}(2022)\citenamefont
  {De~Palma}, \citenamefont {Marvian}, \citenamefont {Rouzé},\ and\
  \citenamefont {França}}]{DePalma2022}%
  \BibitemOpen
  \bibfield  {author} {\bibinfo {author} {\bibfnamefont {G.}~\bibnamefont
  {De~Palma}}, \bibinfo {author} {\bibfnamefont {M.}~\bibnamefont {Marvian}},
  \bibinfo {author} {\bibfnamefont {C.}~\bibnamefont {Rouzé}}, \ and\ \bibinfo
  {author} {\bibfnamefont {D.~S.}\ \bibnamefont {França}},\ }\href {\doibase
  10.1103/prxquantum.4.010309} {\bibfield  {journal} {\bibinfo  {journal} {PRX
  Quantum 4 (1), 010309, 2023}\ }\textbf {\bibinfo {volume} {4}},\ \bibinfo
  {pages} {010309} (\bibinfo {year} {2022})},\ \Eprint
  {http://arxiv.org/abs/2204.03455} {arXiv:2204.03455 [quant-ph]} \BibitemShut
  {NoStop}%
\bibitem [{\citenamefont {González-García}\ \emph {et~al.}(2022)\citenamefont
  {González-García}, \citenamefont {Trivedi},\ and\ \citenamefont
  {Cirac}}]{Gonzlez_Garca_2022}%
  \BibitemOpen
  \bibfield  {author} {\bibinfo {author} {\bibfnamefont {G.}~\bibnamefont
  {González-García}}, \bibinfo {author} {\bibfnamefont {R.}~\bibnamefont
  {Trivedi}}, \ and\ \bibinfo {author} {\bibfnamefont {J.~I.}\ \bibnamefont
  {Cirac}},\ }\href {\doibase 10.1103/prxquantum.3.040326} {\bibfield
  {journal} {\bibinfo  {journal} {PRX Quantum}\ }\textbf {\bibinfo {volume}
  {3}} (\bibinfo {year} {2022}),\ 10.1103/prxquantum.3.040326}\BibitemShut
  {NoStop}%
\bibitem [{\citenamefont {Brady}\ and\ \citenamefont
  {Hadfield}(2024)}]{brady2024}%
  \BibitemOpen
  \bibfield  {author} {\bibinfo {author} {\bibfnamefont {L.~T.}\ \bibnamefont
  {Brady}}\ and\ \bibinfo {author} {\bibfnamefont {S.}~\bibnamefont
  {Hadfield}},\ }\href {\doibase 10.1103/physreva.110.052435} {\bibfield
  {journal} {\bibinfo  {journal} {Physical Review A}\ }\textbf {\bibinfo
  {volume} {110}} (\bibinfo {year} {2024}),\
  10.1103/physreva.110.052435}\BibitemShut {NoStop}%
\bibitem [{\citenamefont {Dupont}\ and\ \citenamefont
  {Sundar}(2024)}]{Dupont2024}%
  \BibitemOpen
  \bibfield  {author} {\bibinfo {author} {\bibfnamefont {M.}~\bibnamefont
  {Dupont}}\ and\ \bibinfo {author} {\bibfnamefont {B.}~\bibnamefont
  {Sundar}},\ }\href {\doibase 10.1103/physreva.109.012429} {\bibfield
  {journal} {\bibinfo  {journal} {Physical Review A}\ }\textbf {\bibinfo
  {volume} {109}} (\bibinfo {year} {2024}),\
  10.1103/physreva.109.012429}\BibitemShut {NoStop}%
\bibitem [{\citenamefont {Dupont}\ \emph {et~al.}(2023)\citenamefont {Dupont},
  \citenamefont {Evert}, \citenamefont {Hodson}, \citenamefont {Sundar},
  \citenamefont {Jeffrey}, \citenamefont {Yamaguchi}, \citenamefont {Feng},
  \citenamefont {Maciejewski}, \citenamefont {Hadfield}, \citenamefont {Alam},
  \citenamefont {Wang}, \citenamefont {Grabbe}, \citenamefont {Lott},
  \citenamefont {Rieffel}, \citenamefont {Venturelli},\ and\ \citenamefont
  {Reagor}}]{Dupont2023a}%
  \BibitemOpen
  \bibfield  {author} {\bibinfo {author} {\bibfnamefont {M.}~\bibnamefont
  {Dupont}}, \bibinfo {author} {\bibfnamefont {B.}~\bibnamefont {Evert}},
  \bibinfo {author} {\bibfnamefont {M.~J.}\ \bibnamefont {Hodson}}, \bibinfo
  {author} {\bibfnamefont {B.}~\bibnamefont {Sundar}}, \bibinfo {author}
  {\bibfnamefont {S.}~\bibnamefont {Jeffrey}}, \bibinfo {author} {\bibfnamefont
  {Y.}~\bibnamefont {Yamaguchi}}, \bibinfo {author} {\bibfnamefont
  {D.}~\bibnamefont {Feng}}, \bibinfo {author} {\bibfnamefont {F.~B.}\
  \bibnamefont {Maciejewski}}, \bibinfo {author} {\bibfnamefont
  {S.}~\bibnamefont {Hadfield}}, \bibinfo {author} {\bibfnamefont {M.~S.}\
  \bibnamefont {Alam}}, \bibinfo {author} {\bibfnamefont {Z.}~\bibnamefont
  {Wang}}, \bibinfo {author} {\bibfnamefont {S.}~\bibnamefont {Grabbe}},
  \bibinfo {author} {\bibfnamefont {P.~A.}\ \bibnamefont {Lott}}, \bibinfo
  {author} {\bibfnamefont {E.~G.}\ \bibnamefont {Rieffel}}, \bibinfo {author}
  {\bibfnamefont {D.}~\bibnamefont {Venturelli}}, \ and\ \bibinfo {author}
  {\bibfnamefont {M.~J.}\ \bibnamefont {Reagor}},\ }\href {\doibase
  10.1126/sciadv.adi0487} {\bibfield  {journal} {\bibinfo  {journal} {Science
  Advances 9, 45 (2023)}\ }\textbf {\bibinfo {volume} {9}} (\bibinfo {year}
  {2023}),\ 10.1126/sciadv.adi0487},\ \Eprint {http://arxiv.org/abs/2303.05509}
  {arXiv:2303.05509 [quant-ph]} \BibitemShut {NoStop}%
\bibitem [{\citenamefont {Finžgar}\ \emph {et~al.}(2024)\citenamefont
  {Finžgar}, \citenamefont {Kerschbaumer}, \citenamefont {Schuetz},
  \citenamefont {Mendl},\ and\ \citenamefont {Katzgraber}}]{Finzgar2024}%
  \BibitemOpen
  \bibfield  {author} {\bibinfo {author} {\bibfnamefont {J.~R.}\ \bibnamefont
  {Finžgar}}, \bibinfo {author} {\bibfnamefont {A.}~\bibnamefont
  {Kerschbaumer}}, \bibinfo {author} {\bibfnamefont {M.~J.}\ \bibnamefont
  {Schuetz}}, \bibinfo {author} {\bibfnamefont {C.~B.}\ \bibnamefont {Mendl}},
  \ and\ \bibinfo {author} {\bibfnamefont {H.~G.}\ \bibnamefont {Katzgraber}},\
  }\href {\doibase 10.1103/prxquantum.5.020327} {\bibfield  {journal} {\bibinfo
   {journal} {PRX Quantum}\ }\textbf {\bibinfo {volume} {5}} (\bibinfo {year}
  {2024}),\ 10.1103/prxquantum.5.020327}\BibitemShut {NoStop}%
\bibitem [{\citenamefont {Frieze}(1990)}]{Frieze1990}%
  \BibitemOpen
  \bibfield  {author} {\bibinfo {author} {\bibfnamefont {A.}~\bibnamefont
  {Frieze}},\ }\href {\doibase 10.1016/0012-365X(90)90136-R} {\bibfield
  {journal} {\bibinfo  {journal} {Discrete Mathematics}\ }\textbf {\bibinfo
  {volume} {81}},\ \bibinfo {pages} {171} (\bibinfo {year} {1990})}\BibitemShut
  {NoStop}%
\bibitem [{\citenamefont {Frieze}\ and\ \citenamefont
  {Łuczak}(1992)}]{Frieze1992}%
  \BibitemOpen
  \bibfield  {author} {\bibinfo {author} {\bibfnamefont {A.}~\bibnamefont
  {Frieze}}\ and\ \bibinfo {author} {\bibfnamefont {T.}~\bibnamefont
  {Łuczak}},\ }\href {\doibase https://doi.org/10.1016/0095-8956(92)90070-E}
  {\bibfield  {journal} {\bibinfo  {journal} {Journal of Combinatorial Theory,
  Series B}\ }\textbf {\bibinfo {volume} {54}},\ \bibinfo {pages} {123}
  (\bibinfo {year} {1992})}\BibitemShut {NoStop}%
\bibitem [{\citenamefont {Boppana}\ and\ \citenamefont
  {H{\aa}ld{\'o}rsson}(1992)}]{Halldorsson1992}%
  \BibitemOpen
  \bibfield  {author} {\bibinfo {author} {\bibfnamefont {R.~B.}\ \bibnamefont
  {Boppana}}\ and\ \bibinfo {author} {\bibfnamefont {M.~M.}\ \bibnamefont
  {H{\aa}ld{\'o}rsson}},\ }\href {\doibase 10.1007/BF01994876} {\bibfield
  {journal} {\bibinfo  {journal} {BIT Numerical Mathematics}\ }\textbf
  {\bibinfo {volume} {32}},\ \bibinfo {pages} {180} (\bibinfo {year}
  {1992})}\BibitemShut {NoStop}%
\bibitem [{\citenamefont {H{\aa}ld{\'o}rsson}(1997)}]{Halldorsson1998}%
  \BibitemOpen
  \bibfield  {author} {\bibinfo {author} {\bibfnamefont {M.~M.}\ \bibnamefont
  {H{\aa}ld{\'o}rsson}},\ }\href {\doibase 10.1006/jagm.1997.0898} {\bibfield
  {journal} {\bibinfo  {journal} {Journal of Algorithms}\ }\textbf {\bibinfo
  {volume} {25}},\ \bibinfo {pages} {1} (\bibinfo {year} {1997})}\BibitemShut
  {NoStop}%
\bibitem [{\citenamefont {Wormald}(1995)}]{Wormald1999}%
  \BibitemOpen
  \bibfield  {author} {\bibinfo {author} {\bibfnamefont {N.~C.}\ \bibnamefont
  {Wormald}},\ }\href {\doibase 10.1214/aoap/1177004824} {\bibfield  {journal}
  {\bibinfo  {journal} {Annals of Applied Probability}\ }\textbf {\bibinfo
  {volume} {5}},\ \bibinfo {pages} {1217} (\bibinfo {year} {1995})}\BibitemShut
  {NoStop}%
\bibitem [{\citenamefont {Cerezo}\ \emph {et~al.}(2021)\citenamefont {Cerezo},
  \citenamefont {Sone}, \citenamefont {Volkoff}, \citenamefont {Cincio},\ and\
  \citenamefont {Coles}}]{Cerezo2021}%
  \BibitemOpen
  \bibfield  {author} {\bibinfo {author} {\bibfnamefont {M.}~\bibnamefont
  {Cerezo}}, \bibinfo {author} {\bibfnamefont {A.}~\bibnamefont {Sone}},
  \bibinfo {author} {\bibfnamefont {T.}~\bibnamefont {Volkoff}}, \bibinfo
  {author} {\bibfnamefont {L.}~\bibnamefont {Cincio}}, \ and\ \bibinfo {author}
  {\bibfnamefont {P.~J.}\ \bibnamefont {Coles}},\ }\href {\doibase
  10.1038/s41467-021-21728-w} {\bibfield  {journal} {\bibinfo  {journal}
  {Nature Communications}\ }\textbf {\bibinfo {volume} {12}},\ \bibinfo {pages}
  {1791} (\bibinfo {year} {2021})}\BibitemShut {NoStop}%
\bibitem [{\citenamefont {Larocca}\ \emph {et~al.}(2025)\citenamefont
  {Larocca}, \citenamefont {Thanasilp}, \citenamefont {Wang}, \citenamefont
  {Sharma}, \citenamefont {Biamonte}, \citenamefont {Coles}, \citenamefont
  {Cincio}, \citenamefont {McClean}, \citenamefont {Holmes},\ and\
  \citenamefont {Cerezo}}]{Larocca2025}%
  \BibitemOpen
  \bibfield  {author} {\bibinfo {author} {\bibfnamefont {M.}~\bibnamefont
  {Larocca}}, \bibinfo {author} {\bibfnamefont {S.}~\bibnamefont {Thanasilp}},
  \bibinfo {author} {\bibfnamefont {S.}~\bibnamefont {Wang}}, \bibinfo {author}
  {\bibfnamefont {K.}~\bibnamefont {Sharma}}, \bibinfo {author} {\bibfnamefont
  {J.}~\bibnamefont {Biamonte}}, \bibinfo {author} {\bibfnamefont {P.~J.}\
  \bibnamefont {Coles}}, \bibinfo {author} {\bibfnamefont {L.}~\bibnamefont
  {Cincio}}, \bibinfo {author} {\bibfnamefont {J.~R.}\ \bibnamefont {McClean}},
  \bibinfo {author} {\bibfnamefont {Z.}~\bibnamefont {Holmes}}, \ and\ \bibinfo
  {author} {\bibfnamefont {M.}~\bibnamefont {Cerezo}},\ }\href {\doibase
  10.1038/s42254-025-00813-9} {\bibfield  {journal} {\bibinfo  {journal}
  {Nature Reviews Physics}\ }\textbf {\bibinfo {volume} {7}},\ \bibinfo {pages}
  {174–189} (\bibinfo {year} {2025})}\BibitemShut {NoStop}%
\bibitem [{\citenamefont {Streif}\ and\ \citenamefont
  {Leib}(2020)}]{Streif2020}%
  \BibitemOpen
  \bibfield  {author} {\bibinfo {author} {\bibfnamefont {M.}~\bibnamefont
  {Streif}}\ and\ \bibinfo {author} {\bibfnamefont {M.}~\bibnamefont {Leib}},\
  }\href {\doibase 10.1088/2058-9565/ab8c2b} {\bibfield  {journal} {\bibinfo
  {journal} {Quantum Science and Technology}\ }\textbf {\bibinfo {volume}
  {5}},\ \bibinfo {pages} {034008} (\bibinfo {year} {2020})}\BibitemShut
  {NoStop}%
\bibitem [{\citenamefont {Wurtz}\ and\ \citenamefont {Love}(2020)}]{Wurtz2020}%
  \BibitemOpen
  \bibfield  {author} {\bibinfo {author} {\bibfnamefont {J.}~\bibnamefont
  {Wurtz}}\ and\ \bibinfo {author} {\bibfnamefont {P.~J.}\ \bibnamefont
  {Love}},\ }\href {\doibase 10.1103/physreva.103.042612} {\bibfield  {journal}
  {\bibinfo  {journal} {Phys. Rev. A 103, 042612 (2021)}\ }\textbf {\bibinfo
  {volume} {103}},\ \bibinfo {pages} {042612} (\bibinfo {year} {2020})},\
  \Eprint {http://arxiv.org/abs/2010.11209} {arXiv:2010.11209 [quant-ph]}
  \BibitemShut {NoStop}%
\bibitem [{\citenamefont {Wybo}\ and\ \citenamefont {Leib}(2025)}]{Wybo2025}%
  \BibitemOpen
  \bibfield  {author} {\bibinfo {author} {\bibfnamefont {E.}~\bibnamefont
  {Wybo}}\ and\ \bibinfo {author} {\bibfnamefont {M.}~\bibnamefont {Leib}},\
  }\href {\doibase 10.22331/q-2025-10-22-1892} {\bibfield  {journal} {\bibinfo
  {journal} {{Quantum}}\ }\textbf {\bibinfo {volume} {9}},\ \bibinfo {pages}
  {1892} (\bibinfo {year} {2025})}\BibitemShut {NoStop}%
\bibitem [{\citenamefont {Abdurakhimov}\ \emph {et~al.}(2024)\citenamefont
  {Abdurakhimov}, \citenamefont {Adam}, \citenamefont {Ahmad}, \citenamefont
  {Ahonen}, \citenamefont {Algaba}, \citenamefont {Alonso}, \citenamefont
  {Bergholm}, \citenamefont {Beriwal}, \citenamefont {Beuerle}, \citenamefont
  {Bockstiegel} \emph {et~al.}}]{abdurakhimov2024}%
  \BibitemOpen
  \bibfield  {author} {\bibinfo {author} {\bibfnamefont {L.}~\bibnamefont
  {Abdurakhimov}}, \bibinfo {author} {\bibfnamefont {J.}~\bibnamefont {Adam}},
  \bibinfo {author} {\bibfnamefont {H.}~\bibnamefont {Ahmad}}, \bibinfo
  {author} {\bibfnamefont {O.}~\bibnamefont {Ahonen}}, \bibinfo {author}
  {\bibfnamefont {M.}~\bibnamefont {Algaba}}, \bibinfo {author} {\bibfnamefont
  {G.}~\bibnamefont {Alonso}}, \bibinfo {author} {\bibfnamefont
  {V.}~\bibnamefont {Bergholm}}, \bibinfo {author} {\bibfnamefont
  {R.}~\bibnamefont {Beriwal}}, \bibinfo {author} {\bibfnamefont
  {M.}~\bibnamefont {Beuerle}}, \bibinfo {author} {\bibfnamefont
  {C.}~\bibnamefont {Bockstiegel}},  \emph {et~al.},\ }\href@noop {} {\bibfield
   {journal} {\bibinfo  {journal} {arXiv preprint arXiv:2408.12433}\ }
  (\bibinfo {year} {2024})}\BibitemShut {NoStop}%
\bibitem [{\citenamefont {Marino}\ and\ \citenamefont
  {Kirkpatrick}(2020)}]{Marino2020}%
  \BibitemOpen
  \bibfield  {author} {\bibinfo {author} {\bibfnamefont {R.}~\bibnamefont
  {Marino}}\ and\ \bibinfo {author} {\bibfnamefont {S.}~\bibnamefont
  {Kirkpatrick}},\ }\href {\doibase 10.48550/ARXIV.2003.12293} {\  (\bibinfo
  {year} {2020}),\ 10.48550/ARXIV.2003.12293},\ \Eprint
  {http://arxiv.org/abs/2003.12293} {arXiv:2003.12293 [cs.DS]} \BibitemShut
  {NoStop}%
\bibitem [{\citenamefont {Garey}\ and\ \citenamefont
  {Johnson}(1979)}]{GareyJohnson1979}%
  \BibitemOpen
  \bibfield  {author} {\bibinfo {author} {\bibfnamefont {M.~R.}\ \bibnamefont
  {Garey}}\ and\ \bibinfo {author} {\bibfnamefont {D.~S.}\ \bibnamefont
  {Johnson}},\ }\href@noop {} {\emph {\bibinfo {title} {Computers and
  Intractability: A Guide to the Theory of NP-Completeness}}}\ (\bibinfo
  {publisher} {W. H. Freeman},\ \bibinfo {year} {1979})\BibitemShut {NoStop}%
\bibitem [{\citenamefont {H{\aa}stad}(1999)}]{Hastad1999}%
  \BibitemOpen
  \bibfield  {author} {\bibinfo {author} {\bibfnamefont {J.}~\bibnamefont
  {H{\aa}stad}},\ }\href@noop {} {\bibfield  {journal} {\bibinfo  {journal}
  {Acta Mathematica}\ }\textbf {\bibinfo {volume} {182}},\ \bibinfo {pages}
  {105} (\bibinfo {year} {1999})}\BibitemShut {NoStop}%
\bibitem [{\citenamefont {Wurtz}\ \emph {et~al.}(2024)\citenamefont {Wurtz},
  \citenamefont {Lopes}, \citenamefont {Gorgulla}, \citenamefont {Gemelke},
  \citenamefont {Keesling},\ and\ \citenamefont {Wang}}]{Wurtz2024}%
  \BibitemOpen
  \bibfield  {author} {\bibinfo {author} {\bibfnamefont {J.}~\bibnamefont
  {Wurtz}}, \bibinfo {author} {\bibfnamefont {P.~L.~S.}\ \bibnamefont {Lopes}},
  \bibinfo {author} {\bibfnamefont {C.}~\bibnamefont {Gorgulla}}, \bibinfo
  {author} {\bibfnamefont {N.}~\bibnamefont {Gemelke}}, \bibinfo {author}
  {\bibfnamefont {A.}~\bibnamefont {Keesling}}, \ and\ \bibinfo {author}
  {\bibfnamefont {S.}~\bibnamefont {Wang}},\ }\href
  {https://arxiv.org/abs/2205.08500} {\enquote {\bibinfo {title} {Industry
  applications of neutral-atom quantum computing solving independent set
  problems},}\ } (\bibinfo {year} {2024}),\ \Eprint
  {http://arxiv.org/abs/2205.08500} {arXiv:2205.08500 [quant-ph]} \BibitemShut
  {NoStop}%
\bibitem [{\citenamefont {Kieritz}\ \emph {et~al.}(2010)\citenamefont
  {Kieritz}, \citenamefont {Luxen}, \citenamefont {Sanders},\ and\
  \citenamefont {Vetter}}]{Kieritz2010}%
  \BibitemOpen
  \bibfield  {author} {\bibinfo {author} {\bibfnamefont {T.}~\bibnamefont
  {Kieritz}}, \bibinfo {author} {\bibfnamefont {D.}~\bibnamefont {Luxen}},
  \bibinfo {author} {\bibfnamefont {P.}~\bibnamefont {Sanders}}, \ and\
  \bibinfo {author} {\bibfnamefont {C.}~\bibnamefont {Vetter}},\ }in\
  \href@noop {} {\emph {\bibinfo {booktitle} {Experimental Algorithms}}},\
  \bibinfo {editor} {edited by\ \bibinfo {editor} {\bibfnamefont
  {P.}~\bibnamefont {Festa}}}\ (\bibinfo  {publisher} {Springer Berlin
  Heidelberg},\ \bibinfo {address} {Berlin, Heidelberg},\ \bibinfo {year}
  {2010})\ pp.\ \bibinfo {pages} {83--93}\BibitemShut {NoStop}%
\bibitem [{\citenamefont {Sander}\ \emph {et~al.}(2008)\citenamefont {Sander},
  \citenamefont {Nehab}, \citenamefont {Chlamtac},\ and\ \citenamefont
  {Hoppe}}]{Sander2008}%
  \BibitemOpen
  \bibfield  {author} {\bibinfo {author} {\bibfnamefont {P.~V.}\ \bibnamefont
  {Sander}}, \bibinfo {author} {\bibfnamefont {D.}~\bibnamefont {Nehab}},
  \bibinfo {author} {\bibfnamefont {E.}~\bibnamefont {Chlamtac}}, \ and\
  \bibinfo {author} {\bibfnamefont {H.}~\bibnamefont {Hoppe}},\ }\href
  {\doibase 10.1145/1409060.1409097} {\bibfield  {journal} {\bibinfo  {journal}
  {ACM Trans. Graph.}\ }\textbf {\bibinfo {volume} {27}} (\bibinfo {year}
  {2008}),\ 10.1145/1409060.1409097}\BibitemShut {NoStop}%
\bibitem [{\citenamefont {Butenko}\ and\ \citenamefont
  {Wilhelm}(2006)}]{Butenko2006}%
  \BibitemOpen
  \bibfield  {author} {\bibinfo {author} {\bibfnamefont {S.}~\bibnamefont
  {Butenko}}\ and\ \bibinfo {author} {\bibfnamefont {W.}~\bibnamefont
  {Wilhelm}},\ }\href {\doibase https://doi.org/10.1016/j.ejor.2005.05.026}
  {\bibfield  {journal} {\bibinfo  {journal} {European Journal of Operational
  Research}\ }\textbf {\bibinfo {volume} {173}},\ \bibinfo {pages} {1}
  (\bibinfo {year} {2006})}\BibitemShut {NoStop}%
\bibitem [{\citenamefont {Cheng}\ \emph {et~al.}(2008)\citenamefont {Cheng},
  \citenamefont {Lu}, \citenamefont {Vendruscolo}, \citenamefont {Lio’},\
  and\ \citenamefont {Blundell}}]{Cheng2008}%
  \BibitemOpen
  \bibfield  {author} {\bibinfo {author} {\bibfnamefont {T.~M.~K.}\
  \bibnamefont {Cheng}}, \bibinfo {author} {\bibfnamefont {Y.-E.}\ \bibnamefont
  {Lu}}, \bibinfo {author} {\bibfnamefont {M.}~\bibnamefont {Vendruscolo}},
  \bibinfo {author} {\bibfnamefont {P.}~\bibnamefont {Lio’}}, \ and\ \bibinfo
  {author} {\bibfnamefont {T.~L.}\ \bibnamefont {Blundell}},\ }\href {\doibase
  10.1371/journal.pcbi.1000135} {\bibfield  {journal} {\bibinfo  {journal}
  {PLoS Computational Biology}\ }\textbf {\bibinfo {volume} {4}},\ \bibinfo
  {pages} {e1000135} (\bibinfo {year} {2008})}\BibitemShut {NoStop}%
\bibitem [{\citenamefont {Hall\'{o}rsson}\ and\ \citenamefont
  {Radhakrishnan}(1997)}]{Hallorsson1997}%
  \BibitemOpen
  \bibfield  {author} {\bibinfo {author} {\bibfnamefont {M.~M.}\ \bibnamefont
  {Hall\'{o}rsson}}\ and\ \bibinfo {author} {\bibfnamefont {J.}~\bibnamefont
  {Radhakrishnan}},\ }\href {\doibase 10.1007/bf02523693} {\bibfield  {journal}
  {\bibinfo  {journal} {Algorithmica}\ }\textbf {\bibinfo {volume} {18}},\
  \bibinfo {pages} {145–163} (\bibinfo {year} {1997})}\BibitemShut {NoStop}%
\bibitem [{\citenamefont {Csóka}\ \emph {et~al.}(2015)\citenamefont {Csóka},
  \citenamefont {Gerencsér}, \citenamefont {Harangi},\ and\ \citenamefont
  {Virág}}]{Csoka2015}%
  \BibitemOpen
  \bibfield  {author} {\bibinfo {author} {\bibfnamefont {E.}~\bibnamefont
  {Csóka}}, \bibinfo {author} {\bibfnamefont {B.}~\bibnamefont {Gerencsér}},
  \bibinfo {author} {\bibfnamefont {V.}~\bibnamefont {Harangi}}, \ and\
  \bibinfo {author} {\bibfnamefont {B.}~\bibnamefont {Virág}},\ }\href
  {\doibase https://doi.org/10.1002/rsa.20547} {\bibfield  {journal} {\bibinfo
  {journal} {Random Structures \& Algorithms}\ }\textbf {\bibinfo {volume}
  {47}},\ \bibinfo {pages} {284} (\bibinfo {year} {2015})}\BibitemShut
  {NoStop}%
\bibitem [{\citenamefont {Frieze}\ and\ \citenamefont
  {Suen}(1994)}]{Frieze1994}%
  \BibitemOpen
  \bibfield  {author} {\bibinfo {author} {\bibfnamefont {A.}~\bibnamefont
  {Frieze}}\ and\ \bibinfo {author} {\bibfnamefont {S.}~\bibnamefont {Suen}},\
  }\href {\doibase https://doi.org/10.1002/rsa.3240050504} {\bibfield
  {journal} {\bibinfo  {journal} {Random Structures \& Algorithms}\ }\textbf
  {\bibinfo {volume} {5}},\ \bibinfo {pages} {649} (\bibinfo {year}
  {1994})}\BibitemShut {NoStop}%
\bibitem [{\citenamefont {McKay}(1987)}]{McKay1987}%
  \BibitemOpen
  \bibfield  {author} {\bibinfo {author} {\bibfnamefont {B.~D.}\ \bibnamefont
  {McKay}},\ }\href@noop {} {\bibfield  {journal} {\bibinfo  {journal} {Ars
  Combinatoria, 23A 179-185.}\ } (\bibinfo {year} {1987})}\BibitemShut
  {NoStop}%
\bibitem [{\citenamefont {Balogh}\ \emph {et~al.}(2017)\citenamefont {Balogh},
  \citenamefont {Kostochka},\ and\ \citenamefont {Liu}}]{Balogh2017}%
  \BibitemOpen
  \bibfield  {author} {\bibinfo {author} {\bibfnamefont {J.}~\bibnamefont
  {Balogh}}, \bibinfo {author} {\bibfnamefont {A.}~\bibnamefont {Kostochka}}, \
  and\ \bibinfo {author} {\bibfnamefont {X.}~\bibnamefont {Liu}},\ }\href
  {\doibase 10.48550/ARXIV.1708.03996} {\  (\bibinfo {year} {2017}),\
  10.48550/ARXIV.1708.03996},\ \Eprint {http://arxiv.org/abs/1708.03996}
  {arXiv:1708.03996 [math.CO]} \BibitemShut {NoStop}%
\bibitem [{\citenamefont {Ozaeta}\ \emph {et~al.}(2020)\citenamefont {Ozaeta},
  \citenamefont {van Dam},\ and\ \citenamefont {McMahon}}]{Ozaeta2020}%
  \BibitemOpen
  \bibfield  {author} {\bibinfo {author} {\bibfnamefont {A.}~\bibnamefont
  {Ozaeta}}, \bibinfo {author} {\bibfnamefont {W.}~\bibnamefont {van Dam}}, \
  and\ \bibinfo {author} {\bibfnamefont {P.~L.}\ \bibnamefont {McMahon}},\
  }\href {\doibase 10.1088/2058-9565/ac9013} {\bibfield  {journal} {\bibinfo
  {journal} {Quantum Sci. Technol. 7 045036 (2022)}\ }\textbf {\bibinfo
  {volume} {7}},\ \bibinfo {pages} {045036} (\bibinfo {year} {2020})},\ \Eprint
  {http://arxiv.org/abs/2012.03421} {arXiv:2012.03421 [quant-ph]} \BibitemShut
  {NoStop}%
\bibitem [{\citenamefont {Gray}(2018)}]{Gray2018quimb}%
  \BibitemOpen
  \bibfield  {author} {\bibinfo {author} {\bibfnamefont {J.}~\bibnamefont
  {Gray}},\ }\href {\doibase 10.21105/joss.00819} {\bibfield  {journal}
  {\bibinfo  {journal} {Journal of Open Source Software}\ }\textbf {\bibinfo
  {volume} {3}},\ \bibinfo {pages} {819} (\bibinfo {year} {2018})}\BibitemShut
  {NoStop}%
\bibitem [{\citenamefont {Markov}\ and\ \citenamefont {Shi}(2008)}]{Shi2008}%
  \BibitemOpen
  \bibfield  {author} {\bibinfo {author} {\bibfnamefont {I.~L.}\ \bibnamefont
  {Markov}}\ and\ \bibinfo {author} {\bibfnamefont {Y.}~\bibnamefont {Shi}},\
  }\href {\doibase 10.1137/050644756} {\bibfield  {journal} {\bibinfo
  {journal} {SIAM Journal on Computing}\ }\textbf {\bibinfo {volume} {38}},\
  \bibinfo {pages} {963} (\bibinfo {year} {2008})},\ \Eprint
  {http://arxiv.org/abs/https://doi.org/10.1137/050644756}
  {https://doi.org/10.1137/050644756} \BibitemShut {NoStop}%
\bibitem [{\citenamefont {Wood}(2011)}]{Wood2011}%
  \BibitemOpen
  \bibfield  {author} {\bibinfo {author} {\bibfnamefont {D.~R.}\ \bibnamefont
  {Wood}},\ }\href {\doibase 10.1002/jgt.21677} {\  (\bibinfo {year} {2011}),\
  10.1002/jgt.21677},\ \Eprint {http://arxiv.org/abs/arXiv:1105.1586}
  {arXiv:1105.1586} \BibitemShut {NoStop}%
\bibitem [{\citenamefont {Kotil}\ \emph {et~al.}(2023)\citenamefont {Kotil},
  \citenamefont {Simkovic},\ and\ \citenamefont {Leib}}]{Kotil2023}%
  \BibitemOpen
  \bibfield  {author} {\bibinfo {author} {\bibfnamefont {A.}~\bibnamefont
  {Kotil}}, \bibinfo {author} {\bibfnamefont {F.}~\bibnamefont {Simkovic}}, \
  and\ \bibinfo {author} {\bibfnamefont {M.}~\bibnamefont {Leib}},\ }\href
  {https://arxiv.org/abs/2312.15982} {\enquote {\bibinfo {title} {Improved
  qubit routing for qaoa circuits},}\ } (\bibinfo {year} {2023}),\ \Eprint
  {http://arxiv.org/abs/2312.15982} {arXiv:2312.15982 [quant-ph]} \BibitemShut
  {NoStop}%
\bibitem [{\citenamefont {Schnorr}\ and\ \citenamefont
  {Shamir}(1986)}]{Schnorr1986}%
  \BibitemOpen
  \bibfield  {author} {\bibinfo {author} {\bibfnamefont {C.-P.}\ \bibnamefont
  {Schnorr}}\ and\ \bibinfo {author} {\bibfnamefont {A.}~\bibnamefont
  {Shamir}},\ }in\ \href {https://api.semanticscholar.org/CorpusID:7988246}
  {\emph {\bibinfo {booktitle} {Symposium on the Theory of Computing}}}\
  (\bibinfo {year} {1986})\BibitemShut {NoStop}%
\bibitem [{\citenamefont {Steiger}\ \emph {et~al.}(2019)\citenamefont
  {Steiger}, \citenamefont {Häner},\ and\ \citenamefont
  {Troyer}}]{Steiger2019}%
  \BibitemOpen
  \bibfield  {author} {\bibinfo {author} {\bibfnamefont {D.~S.}\ \bibnamefont
  {Steiger}}, \bibinfo {author} {\bibfnamefont {T.}~\bibnamefont {Häner}}, \
  and\ \bibinfo {author} {\bibfnamefont {M.}~\bibnamefont {Troyer}},\ }\href
  {\doibase 10.1016/j.micpro.2019.02.003} {\bibfield  {journal} {\bibinfo
  {journal} {Microprocessors and Microsystems}\ }\textbf {\bibinfo {volume}
  {66}},\ \bibinfo {pages} {81–89} (\bibinfo {year} {2019})}\BibitemShut
  {NoStop}%
\bibitem [{\citenamefont {Hoeffding}(1963)}]{Hoeffding1963}%
  \BibitemOpen
  \bibfield  {author} {\bibinfo {author} {\bibfnamefont {W.}~\bibnamefont
  {Hoeffding}},\ }\href {\doibase 10.1080/01621459.1963.10500830} {\bibfield
  {journal} {\bibinfo  {journal} {Journal of the American Statistical
  Association}\ }\textbf {\bibinfo {volume} {58}},\ \bibinfo {pages} {13–30}
  (\bibinfo {year} {1963})}\BibitemShut {NoStop}%
\bibitem [{\citenamefont {Brylinski}\ and\ \citenamefont
  {Brylinski}(2001)}]{Brylinski2001}%
  \BibitemOpen
  \bibfield  {author} {\bibinfo {author} {\bibfnamefont {J.-L.}\ \bibnamefont
  {Brylinski}}\ and\ \bibinfo {author} {\bibfnamefont {R.}~\bibnamefont
  {Brylinski}},\ }\href@noop {} {\  (\bibinfo {year} {2001})},\ \Eprint
  {http://arxiv.org/abs/quant-ph/0108062} {quant-ph/0108062 [quant-ph]}
  \BibitemShut {NoStop}%
\bibitem [{\citenamefont {Morales}\ \emph {et~al.}(2019)\citenamefont
  {Morales}, \citenamefont {Biamonte},\ and\ \citenamefont
  {Zimborás}}]{Morales2019}%
  \BibitemOpen
  \bibfield  {author} {\bibinfo {author} {\bibfnamefont {M.~E.~S.}\
  \bibnamefont {Morales}}, \bibinfo {author} {\bibfnamefont {J.}~\bibnamefont
  {Biamonte}}, \ and\ \bibinfo {author} {\bibfnamefont {Z.}~\bibnamefont
  {Zimborás}},\ }\href {\doibase 10.1007/s11128-020-02748-9} {\bibfield
  {journal} {\bibinfo  {journal} {Quantum Information Processing 19, 291
  (2020)}\ }\textbf {\bibinfo {volume} {19}} (\bibinfo {year} {2019}),\
  10.1007/s11128-020-02748-9},\ \Eprint {http://arxiv.org/abs/1909.03123}
  {arXiv:1909.03123 [quant-ph]} \BibitemShut {NoStop}%
\bibitem [{\citenamefont {Lloyd}(2018)}]{Lloyd2018}%
  \BibitemOpen
  \bibfield  {author} {\bibinfo {author} {\bibfnamefont {S.}~\bibnamefont
  {Lloyd}},\ }\href {\doibase 10.48550/ARXIV.1812.11075} {\  (\bibinfo {year}
  {2018}),\ 10.48550/ARXIV.1812.11075},\ \Eprint
  {http://arxiv.org/abs/1812.11075} {arXiv:1812.11075 [quant-ph]} \BibitemShut
  {NoStop}%
\bibitem [{\citenamefont {Wang}\ \emph {et~al.}(2025)\citenamefont {Wang},
  \citenamefont {Chen}, \citenamefont {Sun},\ and\ \citenamefont
  {Zhang}}]{Wang2025}%
  \BibitemOpen
  \bibfield  {author} {\bibinfo {author} {\bibfnamefont {J.}~\bibnamefont
  {Wang}}, \bibinfo {author} {\bibfnamefont {S.}~\bibnamefont {Chen}}, \bibinfo
  {author} {\bibfnamefont {X.}~\bibnamefont {Sun}}, \ and\ \bibinfo {author}
  {\bibfnamefont {J.}~\bibnamefont {Zhang}},\ }\href@noop {} {\bibfield
  {journal} {\bibinfo  {journal} {arXiv preprint arXiv:2511.20212}\ } (\bibinfo
  {year} {2025})}\BibitemShut {NoStop}%
\bibitem [{\citenamefont {Wallman}\ and\ \citenamefont
  {Emerson}(2016)}]{wallman2016}%
  \BibitemOpen
  \bibfield  {author} {\bibinfo {author} {\bibfnamefont {J.~J.}\ \bibnamefont
  {Wallman}}\ and\ \bibinfo {author} {\bibfnamefont {J.}~\bibnamefont
  {Emerson}},\ }\href@noop {} {\bibfield  {journal} {\bibinfo  {journal}
  {Physical Review A}\ }\textbf {\bibinfo {volume} {94}},\ \bibinfo {pages}
  {052325} (\bibinfo {year} {2016})}\BibitemShut {NoStop}%
\bibitem [{\citenamefont {Hashim}\ \emph {et~al.}(2020)\citenamefont {Hashim},
  \citenamefont {Naik}, \citenamefont {Morvan}, \citenamefont {Ville},
  \citenamefont {Mitchell}, \citenamefont {Kreikebaum}, \citenamefont {Davis},
  \citenamefont {Smith}, \citenamefont {Iancu}, \citenamefont {O'Brien} \emph
  {et~al.}}]{hashim2020}%
  \BibitemOpen
  \bibfield  {author} {\bibinfo {author} {\bibfnamefont {A.}~\bibnamefont
  {Hashim}}, \bibinfo {author} {\bibfnamefont {R.~K.}\ \bibnamefont {Naik}},
  \bibinfo {author} {\bibfnamefont {A.}~\bibnamefont {Morvan}}, \bibinfo
  {author} {\bibfnamefont {J.-L.}\ \bibnamefont {Ville}}, \bibinfo {author}
  {\bibfnamefont {B.}~\bibnamefont {Mitchell}}, \bibinfo {author}
  {\bibfnamefont {J.~M.}\ \bibnamefont {Kreikebaum}}, \bibinfo {author}
  {\bibfnamefont {M.}~\bibnamefont {Davis}}, \bibinfo {author} {\bibfnamefont
  {E.}~\bibnamefont {Smith}}, \bibinfo {author} {\bibfnamefont
  {C.}~\bibnamefont {Iancu}}, \bibinfo {author} {\bibfnamefont {K.~P.}\
  \bibnamefont {O'Brien}},  \emph {et~al.},\ }\href@noop {} {\bibfield
  {journal} {\bibinfo  {journal} {arXiv preprint arXiv:2010.00215}\ } (\bibinfo
  {year} {2020})}\BibitemShut {NoStop}%
\bibitem [{\citenamefont {Funcke}\ \emph {et~al.}(2022)\citenamefont {Funcke},
  \citenamefont {Hartung}, \citenamefont {Jansen}, \citenamefont {K\"{u}hn},
  \citenamefont {Stornati},\ and\ \citenamefont {Wang}}]{Funcke2022}%
  \BibitemOpen
  \bibfield  {author} {\bibinfo {author} {\bibfnamefont {L.}~\bibnamefont
  {Funcke}}, \bibinfo {author} {\bibfnamefont {T.}~\bibnamefont {Hartung}},
  \bibinfo {author} {\bibfnamefont {K.}~\bibnamefont {Jansen}}, \bibinfo
  {author} {\bibfnamefont {S.}~\bibnamefont {K\"{u}hn}}, \bibinfo {author}
  {\bibfnamefont {P.}~\bibnamefont {Stornati}}, \ and\ \bibinfo {author}
  {\bibfnamefont {X.}~\bibnamefont {Wang}},\ }\href {\doibase
  10.1103/physreva.105.062404} {\bibfield  {journal} {\bibinfo  {journal}
  {Physical Review A}\ }\textbf {\bibinfo {volume} {105}} (\bibinfo {year}
  {2022}),\ 10.1103/physreva.105.062404}\BibitemShut {NoStop}%
\bibitem [{\citenamefont {Kirkpatrick}\ \emph {et~al.}(1983)\citenamefont
  {Kirkpatrick}, \citenamefont {Gelatt},\ and\ \citenamefont
  {Vecchi}}]{Kirkpatrick1983}%
  \BibitemOpen
  \bibfield  {author} {\bibinfo {author} {\bibfnamefont {S.}~\bibnamefont
  {Kirkpatrick}}, \bibinfo {author} {\bibfnamefont {C.~D.}\ \bibnamefont
  {Gelatt}}, \ and\ \bibinfo {author} {\bibfnamefont {M.~P.}\ \bibnamefont
  {Vecchi}},\ }\href@noop {} {\bibfield  {journal} {\bibinfo  {journal}
  {Science}\ }\textbf {\bibinfo {volume} {220}},\ \bibinfo {pages} {671}
  (\bibinfo {year} {1983})}\BibitemShut {NoStop}%
\bibitem [{\citenamefont {Angelini}\ and\ \citenamefont
  {Ricci-Tersenghi}(2019)}]{Angelini2019}%
  \BibitemOpen
  \bibfield  {author} {\bibinfo {author} {\bibfnamefont {M.~C.}\ \bibnamefont
  {Angelini}}\ and\ \bibinfo {author} {\bibfnamefont {F.}~\bibnamefont
  {Ricci-Tersenghi}},\ }\href {\doibase 10.1103/PhysRevE.100.013302} {\bibfield
   {journal} {\bibinfo  {journal} {Phys. Rev. E}\ }\textbf {\bibinfo {volume}
  {100}},\ \bibinfo {pages} {013302} (\bibinfo {year} {2019})}\BibitemShut
  {NoStop}%
\bibitem [{\citenamefont {Schuetz}\ \emph {et~al.}(2025)\citenamefont
  {Schuetz}, \citenamefont {Yalovetzky}, \citenamefont {Andrist}, \citenamefont
  {Salton}, \citenamefont {Sun}, \citenamefont {Raymond}, \citenamefont
  {Chakrabarti}, \citenamefont {Acharya}, \citenamefont {Shaydulin},
  \citenamefont {Pistoia},\ and\ \citenamefont {Katzgraber}}]{Schuetz2025}%
  \BibitemOpen
  \bibfield  {author} {\bibinfo {author} {\bibfnamefont {M.~J.~A.}\
  \bibnamefont {Schuetz}}, \bibinfo {author} {\bibfnamefont {R.}~\bibnamefont
  {Yalovetzky}}, \bibinfo {author} {\bibfnamefont {R.~S.}\ \bibnamefont
  {Andrist}}, \bibinfo {author} {\bibfnamefont {G.}~\bibnamefont {Salton}},
  \bibinfo {author} {\bibfnamefont {Y.}~\bibnamefont {Sun}}, \bibinfo {author}
  {\bibfnamefont {R.}~\bibnamefont {Raymond}}, \bibinfo {author} {\bibfnamefont
  {S.}~\bibnamefont {Chakrabarti}}, \bibinfo {author} {\bibfnamefont
  {A.}~\bibnamefont {Acharya}}, \bibinfo {author} {\bibfnamefont
  {R.}~\bibnamefont {Shaydulin}}, \bibinfo {author} {\bibfnamefont
  {M.}~\bibnamefont {Pistoia}}, \ and\ \bibinfo {author} {\bibfnamefont
  {H.~G.}\ \bibnamefont {Katzgraber}},\ }\href
  {https://arxiv.org/abs/2503.12551} {\enquote {\bibinfo {title} {qredumis: A
  quantum-informed reduction algorithm for the maximum independent set
  problem},}\ } (\bibinfo {year} {2025}),\ \Eprint
  {http://arxiv.org/abs/2503.12551} {arXiv:2503.12551 [quant-ph]} \BibitemShut
  {NoStop}%
\end{thebibliography}%


\end{document}